\documentclass[a4paper,fleqn,usenatbib]{mnras}

\usepackage{newtxtext,newtxmath}

\usepackage[T1]{fontenc}
\usepackage{ae,aecompl}

\usepackage{graphicx}	
\usepackage{amsmath}	
\usepackage{amssymb}	

\newcommand{\mbh}{\ensuremath{M_{\rm{BH}}}}

\newcommand{\mst}{\ensuremath{M_\ast}\,}
\newcommand{\er}{\ensuremath{\lambda_{\rm Edd}}}

\newcommand{\lir}{\ensuremath{L_{\rm IR}}}

\newcommand{\wi}{4.5cm}

\newcommand{\rev}[1]{{#1}}

\title[The star formation rate distribution of the most luminous quasars]{No signs of star formation being regulated in the most luminous quasars at $z\sim2$ with ALMA}

\author[Schulze et al.]{Andreas Schulze$^{1,11}$\thanks{E-mail: andreas.schulze@nao.ac.jp}, John D. Silverman$^{2,3}$, Emanuele Daddi$^{4}$, Wiphu Rujopakarn$^{5,6}$, \newauthor 
Daizhong Liu$^{7}$, Malte Schramm$^{1}$,  Vincenzo Mainieri$^{8}$, Masatoshi Imanishi$^{1}$,  \newauthor Michaela Hirschmann$^{9,10}$, Knud Jahnke$^{7}$
\smallskip \\
$^{1}$National Astronomical Observatory of Japan, Mitaka, Tokyo 181-8588, Japan\\
$^{2}$Kavli Institute for the Physics and Mathematics of the Universe (WPI), The University of Tokyo, Kashiwa, Chiba 277-8583, Japan\\
$^{3}$Department of Astronomy, School of Science, The University of Tokyo, 7-3-1 Hongo, Bunkyo, Tokyo 113-0033, Japan\\
$^{4}$CEA, IRFU, DAp, AIM, Universit\'{e} Paris-Saclay, Universit\'{e} Paris Diderot, Sorbonne Paris Cit\'{e}, CNRS, F-91191 Gif-sur-Yvette, France\\
$^{5}$Department of Physics, Faculty of Science, Chulalongkorn University, 254 Phayathai Road, Pathumwan, Bangkok 10330, Thailand\\
$^{6}$National Astronomical Research Institute of Thailand (Public Organization), Don Kaeo, Mae Rim, Chiang Mai 50180, Thailand\\
$^{7}$Max-Planck-Institut f\"ur Astronomie, K\"onigstuhl 17, D-69117, Heidelberg, Germany\\
$^{8}$ESO, Karl-Schwarschild-Strasse 2, 85748, Garching bei M\"unchen, Germany \\
$^{9}$ DARK, Niels Bohr Institute, University of Copenhagen, Lyngbyvej 2, DK-2100 Copenhagen, Denmark \\
$^{10}$Sorbonne Universit\'{e}s, UPMC-CNRS, UMR7095, Institut d' Astrophysique de Paris, F-75014 Paris, France\\
$^{11}$EACOA Fellow
}

\date{Accepted XXX. Received YYY; in original form ZZZ}

\pubyear{2019}

\begin{document}
\label{firstpage}
\pagerange{\pageref{firstpage}--\pageref{lastpage}}
\maketitle

\begin{abstract}

We present ALMA Band~7 observations at $850\mu$m of 20 luminous ($\log\, L_{\rm bol}>46.9$ [erg s$^{-1}$]) unobscured quasars at $z\sim2$. We detect continuum emission for 19/20 quasars. After subtracting an AGN contribution, we measure the total far-IR luminosity for 18 quasars, assuming a  modified blackbody model, and attribute the emission as indicative of the star formation rate (SFR). Our sample can be characterized with a log-normal SFR distribution having a mean of 140 $M_\odot$ yr$^{-1}$ and a dispersion of 0.5~dex. Based on an inference of their stellar masses, the SFRs are similar, in both the mean and dispersion, with star-forming main-sequence galaxies at the equivalent epoch. Thus, there is no evidence for a systematic enhancement or suppression (i.e., regulation or quenching) of star formation in the hosts of the most luminous quasars at $z\sim2$. These results are consistent with the Magneticum cosmological simulation, while in disagreement with a widely recognized phenomenological model that predicts higher SFRs than observed here based on the high bolometric luminosities of this sample. Furthermore, there is only a weak relation between SFR and accretion rate onto their supermassive black holes both for average and individual measurements. We interpret these results as indicative of star formation and quasar accretion being fed from the available gas reservoir(s) in their host with a disconnect due to their different physical sizes, temporal scales, and means of gas processing.
\end{abstract}




\section{Introduction}
It is well established that in an average sense the growth of supermassive black holes (SMBHs) is closely related to the evolution of galaxies and the build-up of their stellar component. This is demonstrated by the tight relation between the SMBH mass and bulge stellar velocity dispersion or bulge mass in the local Universe \citep[e.g.][]{Gebhardt:2000, Haering:2004,Kormendy:2013}, the close match between the  cosmic evolution of the black hole accretion rate density and star formation rate density \citep[e.g.][]{Boyle:1998, Marconi:2004,Silverman:2008,Mullaney:2012b} and by the correlation between average SMBH and stellar growth in star-forming galaxies \citep[e.g.][]{Chen:2013,Delvecchio:2015,Lanzuisi:2017}. Understanding the interplay between SMBH accretion and star formation is essential for our picture of galaxy formation, particularly since feedback effects due to quasars are likely in play.

Theoretical models of galaxy evolution suggest star formation and black hole growth to be linked via a common supply of cold/molecular gas and triggered via major mergers \citep[e.g.][]{DiMatteo:2005, Somerville:2008,Hirschmann:2014}. These models generally require strong AGN feedback which self regulates black hole growth and quenches star formation (SF) in massive galaxies \citep{Croton:2006, Bower:2006,Fabian:2012}.  AGN winds and outflows are promising feedback mechanisms \citep{King:2003,Zubovas:2012,Faucher-Giguere:2012}. Such outflows have been observed in recent years in ionized gas \citep{Rupke:2011,Cano-Diaz:2012,Harrison:2014,Brusa:2015, Carniani:2015, Bischetti:2017} and molecular gas \citep{Feruglio:2010,Cicone:2014,Feruglio:2017,Brusa:2018,Fluetsch:2018}. However, the demographics and the impact of such outflows are still not well understood; thus the importance and details of AGN feedback remains an open issue.  

The most luminous quasars ($L_{\rm bol}>10^{46}$ erg s$^{-1}$) should be particularly effective at impacting the ISM hence star formation \citep{Menci:2008,Zubovas:2012,Hopkins:2016,Bongiorno:2016}, thus it might be expected that a significant fraction of this population is undergoing quenching or have recently been quenched, leading to SFRs below the MS. On the other side, intense episodes of SF and of black hole growth would be fed by the same gas reservoir and potentially triggered by major mergers, so extremely luminous AGN activity would coincide with intense SF in their host galaxies \citep{AnglesAlcazar:2017}, either in a model where the effect of AGN feedback is slow or delayed or in a scenario without AGN feedback. At least for some very luminous AGN such intense SF is observed \citep[e.g.][]{Lutz:2008,Netzer:2016,Pitchford:2016,Banerji:2017,Duras:2017}. A further possibility is that there is no fundamental correlation between AGN and SF activity even for extreme AGN luminosities, implying very luminous AGN would have SFRs consistent with the SF main sequence. In principle also a mix of these scenarios is possible, which would lead to a broad distribution of SFR. 

Several studies have investigated the SF properties of luminous AGN at $z>1$ using {\it Herschel} far-IR photometry \citep{Serjeant:2010,Bonfield:2011,CaoOrjales:2012,Netzer:2014,Khan-Ali:2015,Ma:2015,Netzer:2016,Pitchford:2016,Harris:2016,Dong:2016,Duras:2017,Kalfountzou:2017,Stanley:2017}, {\it Spitzer}/IRS \citep{Lutz:2008} or sub-mm observations \citep{Priddey:2003, Omont:2003,Stevens:2005,Lonsdale:2015,Hatziminaoglou:2018}. Based on individual detections, they typically find high SFRs, sometimes exceeding 1000~$M_\odot$/yr. However, due to the modest SFR sensitivities achieved and the large fraction of non-detections in these observations, these sources are likely biased and do not represent the typical population. While they demonstrate that intense star formation can exist in the hosts of luminous quasars, they do not provide information on the \textit{intrinsic SFR-distribution} in this luminosity regime. For their non-detection these studies have to rely on stacking. However, this approach can be significantly biased if the intrinsic FIR luminosity distribution is highly skewed, so linear means are dominated by a few high-luminosity galaxies. In fact, recent work based on deep 850$\mu$m observations with ALMA \citep{Mullaney:2015,Scholtz:2018} and  SCUBA-2 \citep{Barger:2015} found intrinsic SFR distributions that are consistent with a log-normal rather than a normal distribution. For their sample of moderate-luminosity AGN hosts they found a significant fraction (up to 50\%)  having SFRs below the MS, suggesting different SFR-distributions between star forming galaxies and moderate luminosity AGN, while their linear means are still consistent. This might be an indication for the suppression of SF due to AGN feedback. i.e. quenching. Based on the comparison of a model with and without AGN feedback in the EAGLE hydrodynamical cosmological simulation \citep{Schaye:2015}, recently \citet{Scholtz:2018} argued that a broad width of the intrinsic SFR-distribution, in particular at high stellar mass ($>2\times10^{10} M_\odot$), is a signature of AGN feedback at work on SF in their host galaxies.

Here, we present ALMA observations to establish the intrinsic SFR-distribution of 20 luminous quasars at $z\sim2$, around the peak of AGN and SF activity.  Deriving a SFR for very luminous AGN from short-wavelength continuum is very challenging, due to the significant AGN contribution at almost all wavelengths. In luminous quasars, the AGN contribution is dominant below 20$\mu$m and still significant around 60$\mu$m, which requires a careful spectral energy distribution (SED) decomposition of the AGN and SF component \citep[e.g.][]{Netzer:2016,Duras:2017}. Another major challenge for both \textit{Herschel} imaging and single-dish sub-mm observations is confusion \citep{Hodge:2013,Simpson:2015,Bischetti:2018}. At wavelengths longer than $\sim100\mu$m the AGN SED falls off rapidly \citep[e.g.][]{Deo:2009,Mullaney:2011}, i.e. in the sub-mm regime AGN contamination is minimized, and in most cases negligible. Radio emission will also be contaminated by AGN contribution and thus does not serve as pure SFR tracer \citep{Barger:2015,Zakamska:2016,White:2017}. Therefore, continuum observations in ALMA Band-7 at $850\mu$m observed frame provide a unique, extremely sensitive SFR tracer for very luminous quasars, which does not suffer from source confusion and minimizes contamination from the AGN thus is a cleaner indicator of the SFR in the hosts of rapidly-growing SMBHs.

In Section~\ref{sec:sample} we describe our sample selection, the ALMA observations, archival \textit{Herschel} data  and measurements of SMBH mass and bolometric luminosity.
We present our results on the AGN SED and their star formations rates in Section~\ref{sec:results}.
In  Section~\ref{sec:discu}, we present the star formation rate distribution derived from our sample and compare it to previous observations and theoretical models. Our conclusions are given in Section~\ref{sec:conclu}.
Throughout this paper we use a Hubble constant of $H_0 = 70$ km s$^{-1}$ Mpc$^{-1}$ and cosmological density parameters $\Omega_\mathrm{m} = 0.3$ and $\Omega_\Lambda = 0.7$. We assume a \citet{Chabrier:2003} initial mass function for estimates of stellar mass and star formation rate.

\section{Sample and Observations} \label{sec:sample}
\subsection{Sample selection}
Our study is focused on the extreme luminosity end of optically-selected, unobscured (i.e. broad line/type-1) quasars. Given the orientation scenario to unify obscured and unobscured AGN \citep{Antonucci:1993,Netzer:2015} the restriction to unobscured quasars would a priori not introduce a bias. However, within an evolution framework of black hole activity \citep[e.g.][]{Hopkins:2008,Alexander:2012}, we are specifically targeting AGN after their dust enshrouded blowout phase, in which they emerge as luminous unobscured quasars. We discuss the possible consequences of this selection further in Section~\ref{sec:discu}.

We draw our sample from the Sloan Digital Sky Survey (SDSS) DR7 quasar catalog \citep{Schneider:2010,Shen:2011}, with $\delta<+15$~deg. We focus on a narrow range in redshift  $1.9<z<2.1$ to avoid any uncertainties due to redshift evolution within our sample. This redshift is of special importance since it corresponds to the peak epoch of star formation and AGN activity. Furthermore, it is the highest redshift for which reliable black hole masses based on the broad \ion{Mg}{II} line can be derived from the optical SDSS spectra. We select the most luminous quasars within this redshift range, based on the bolometric luminosity given in \citet{Shen:2011}, $\log L_\mathrm{bol,S11}> 47.3$ [erg s$^{-1}$]. This bolometric luminosity is based on either $L_{3000}$ or $L_{1350}$ using a constant bolometric correction factor of 5.15 and 3.81, respectively \citep{Richards:2006}. \rev{In section~\ref{sec:mbh}, we  present a reevaluation of the bolometric luminosity for our sample. For this we use a different  bolometric correction factor. This choice lowers on average the bolometric correction used throughout the paper by a factor of 1.6, compared to those in \citet{Shen:2011}. Thus, our selection corresponds to $\log L_{\rm bol}>46.9$ [erg s$^{-1}$], for the bolometric correction discussed in section~\ref{sec:mbh}.} These selection criteria result in an initial sample of 62 quasars.

We further removed radio-detected quasars, based on the Faint Images of the Radio Sky at Twenty Centimeter \citep[FIRST;][]{Becker:1995} survey to avoid contamination by AGN synchrotron emission, i.e all our targets have a 1.4~GHz flux below the FIRST detection limit of $\sim1.0$~mJy and are classified as radio-quiet. This cut removed 17 objects, 6 are classified as radio-loud, 6 as radio-quiet but FIRST-detected and 5 quasars are outside of the FIRST footprint.
Out of a total sample of 45 quasars meeting our selection criteria we randomly selected 20 targets for ALMA observations, where we usually grouped two targets together to share a phase calibrator to minimize the overheads. We did not put any restrictions on the sample in respect to the availability of \textit{Herschel} FIR data, in order to be able to draw the rare target population of the most luminous quasars within our redshift window from the full SDSS sky area available to ALMA.  For 5 quasars in the sample \textit{Herschel} \rev{coverage of the objects is}  available in the archive. We further verified that none of our sources are gravitationally lensed \citep{Inada:2012}.

\begin{figure}
\centering
\resizebox{\hsize}{!}{ \includegraphics[clip]{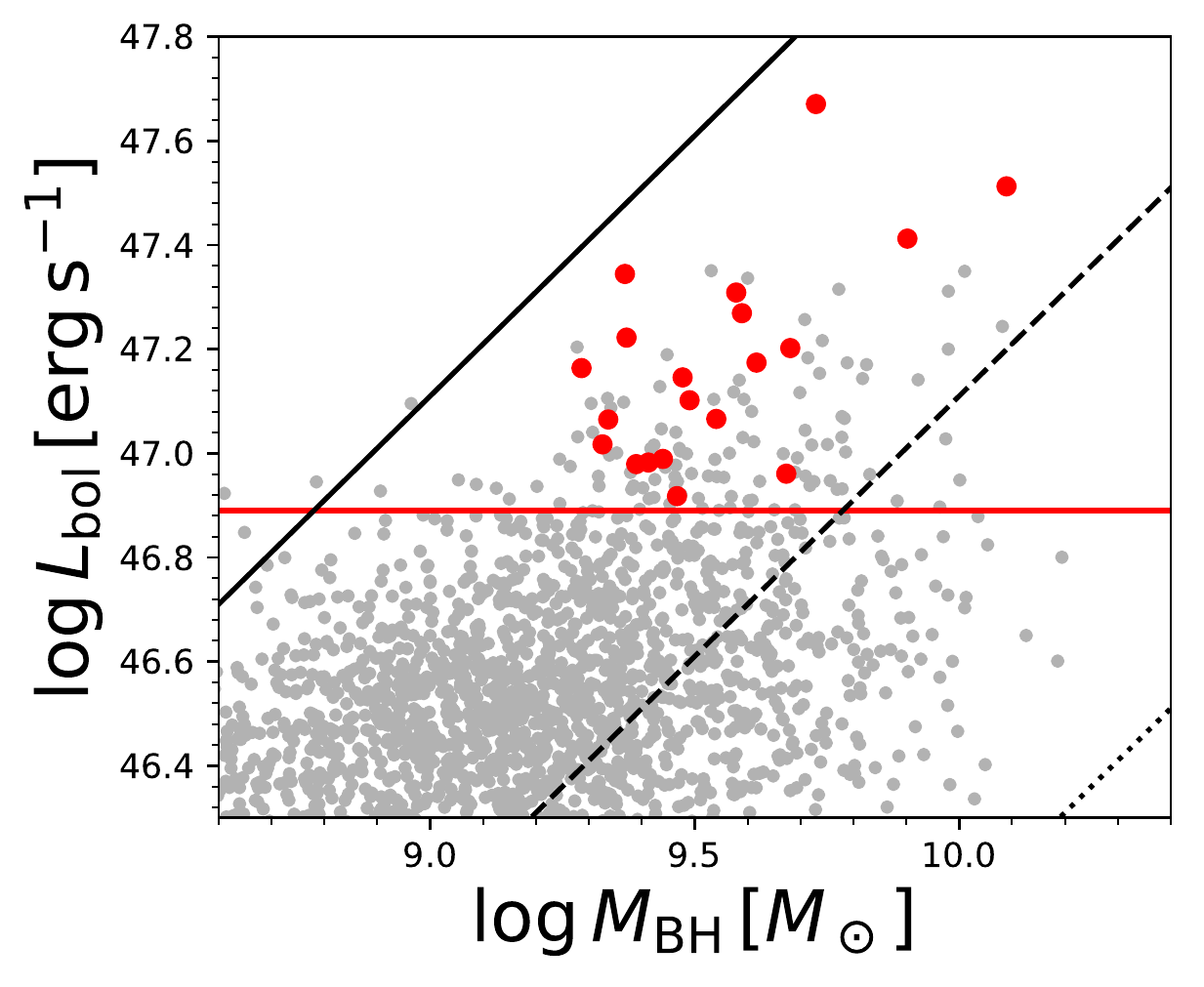} }
\caption{Black hole mass -- luminosity plane of SDSS quasars at redshift $1.9<z<2.1$ and at $\delta<+15\deg$ (gray circles). We highlight our ALMA sample of luminous quasars in red and show our luminosity threshold at $\log L_\mathrm{bol}>46.9$ as red solid line. The solid black, dashed black and dotted black lines indicate Eddington ratios of 1, 0.1 and 0.01.}
\label{fig:sample} 
\end{figure}

\rev{
We show the location of our sample in the SMBH mass-luminosity plane in Figure~\ref{fig:sample} in relation to the general SDSS DR7 quasar population at the same redshift (using black hole masses and bolometric luminosities from section~\ref{sec:mbh}). By our selection, the sample constitutes the most luminous sources in the SDSS DR7 quasar catalog, with $\log L_{\rm bol}>46.9$ [erg s$^{-1}$], black hole masses $\log \mbh>9.2$ [$M_\odot$] and Eddington ratios above 10\%. 
}

\subsection{ALMA Observations}
Our sample of 20 luminous SDSS quasars was observed with ALMA  during Cycle 5 (2017.1.00102.S; PI: A. Schulze) in May 2018 in Band 7. The representative frequency was set to 350.5 GHz (854$\mu$m) with four base bands, each with a band width of 1875~MHz. The high frequency spectral window is chosen to be free of strong atmospheric absorption and to optimize our sensitivity, since the spectral energy distribution of our targets is strongly falling towards lower frequencies. The observations were carried out using 40 antennas in the 12 m array with baselines 15m -- 313m (configuration C43-2). The average major beam size achieved is $\sim 0.9$\arcsec, corresponding to 7.5 kpc at $z=2$ (slightly smaller than our 1$\arcsec$ request).
We did not aim to spatially resolve the emission but rather ensure a measure of the total $850\mu$m continuum emission from the host galaxy. The maximum recoverable scale is $\sim6.5\arcsec$, thus spatially-extended SF emission in the host galaxy is fully recovered.
Our achieved spatial resolution avoids confusion with potentially close companions, in contrast to single dish observations which typically achieve $\gtrsim13\arcsec$ \citep{Dempsey:2013}. 

The requested sensitivity was 0.12 mJy beam$^{-1}$, a level that probes SFRs $\sim$0.2 dex below the MS. For a conservative lower limit on the stellar mass of their host galaxy, $M_*$ = $10^{10.5}\,M_\odot$ typical for luminous quasars at $z\sim2$ \citep{Mechtley:2016}, a galaxy on the MS would have a SFR of $61\,M_\odot$ yr$^{-1}$ based on the relation from \citet{Speagle:2014} at $z=2.0$. We aimed to be sensitive to most galaxies on the MS, thus we set our detection threshold 0.2~dex below this value at $38\,M_\odot$ yr$^{-1}$, corresponding to a FIR luminosity ($8-1000\mu$m) of $\log L_\mathrm{IR}=45.2$ [erg s$^{-1}$] using the relation by \citet{Kennicutt:1998}. This gives a sensitivity of 0.12 mJy beam$^{-1}$ at 345~GHz for a 3$\sigma$ detection. Our achieved sensitivities are always below this value, with a mean around $0.09$~mJy beam$^{-1}$.  

\rev{
To measure fluxes, we processed the ALMA data ourselves by reproducing the observatory calibration with their custom-made script based on Common Astronomy Software Application package \citep[CASA,][]{McMullin:2007}. We converted the data into uvfits format to perform further analysis with the IRAM GILDAS tool working on the uv-space (visibility) data. We measured the 850$\mu$m fluxes and galaxy sizes  by fitting the sources with models directly in the uv space. Gaussian models were used when the emission was found to be resolved at more than 3$\sigma$, while point source models were fit otherwise. Given the large synthesised beam of the ALMA observations and the typical sizes recovered for the resolved galaxies (Table~\ref{tab:res}) we do not expect substantial flux underestimate for galaxies fit with point source models. For more details on the method and for discussions on the advantages of the uv space analysis rather than imaging and cleaning the products we refer to \citet{Valentino:2018} and \citet{Rujopakarn:2019}, respectively.
}
This approach ensures not to loose flux, also for resolved sources, and generally returns the highest SNR flux measurement from the data, contrary to e.g. using aperture photometry. \rev{The typical flux error for our sample is $\sim0.1$~mJy, with the values for each object reported together with the flux values in Table~\ref{tab:res}.}

\begin{figure*}
\centering
\setlength{\unitlength}{1mm}
\begin{picture}(180,185)
\put(0,148){\includegraphics[width=\wi]{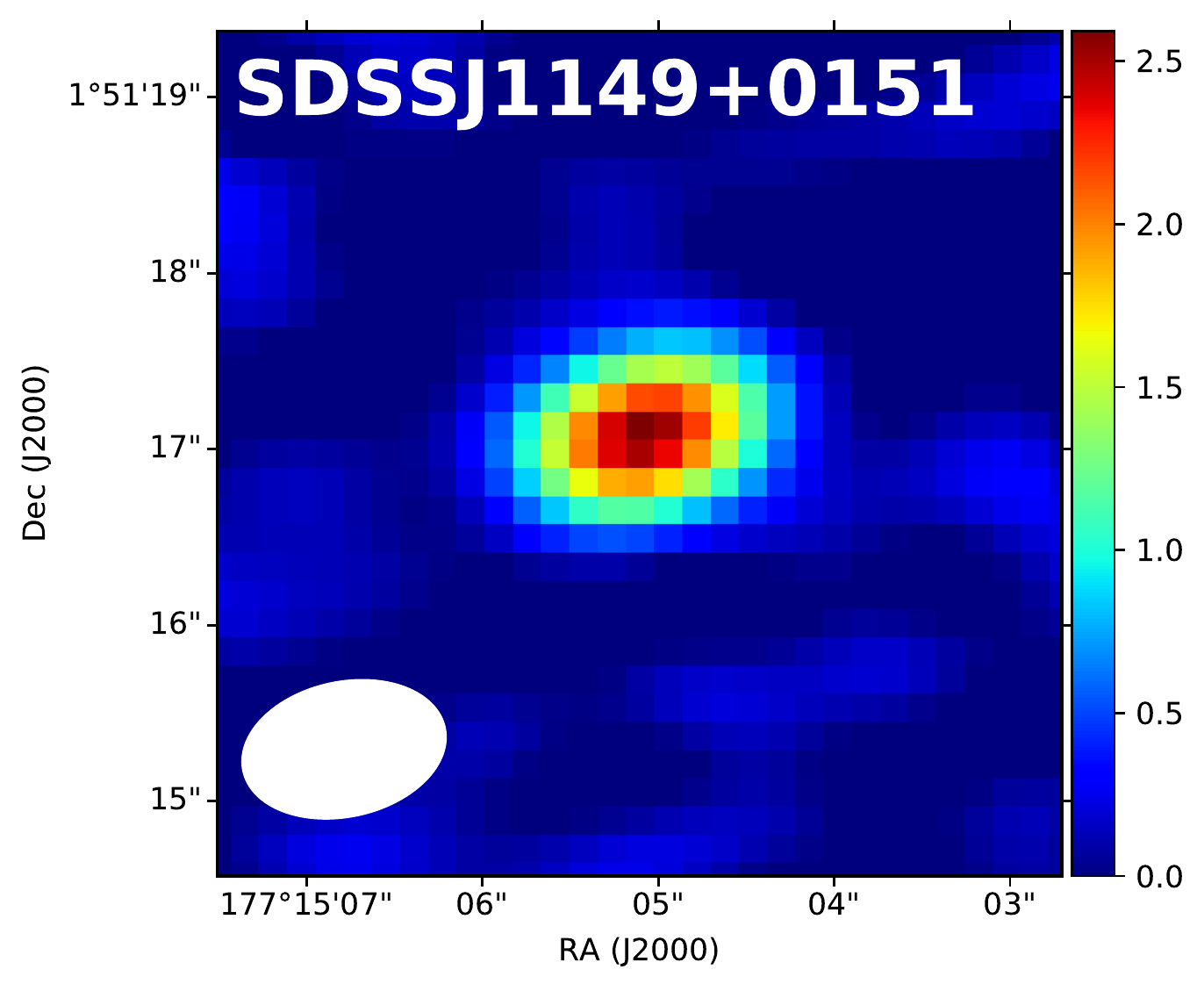}}
\put(45,148){\includegraphics[width=\wi]{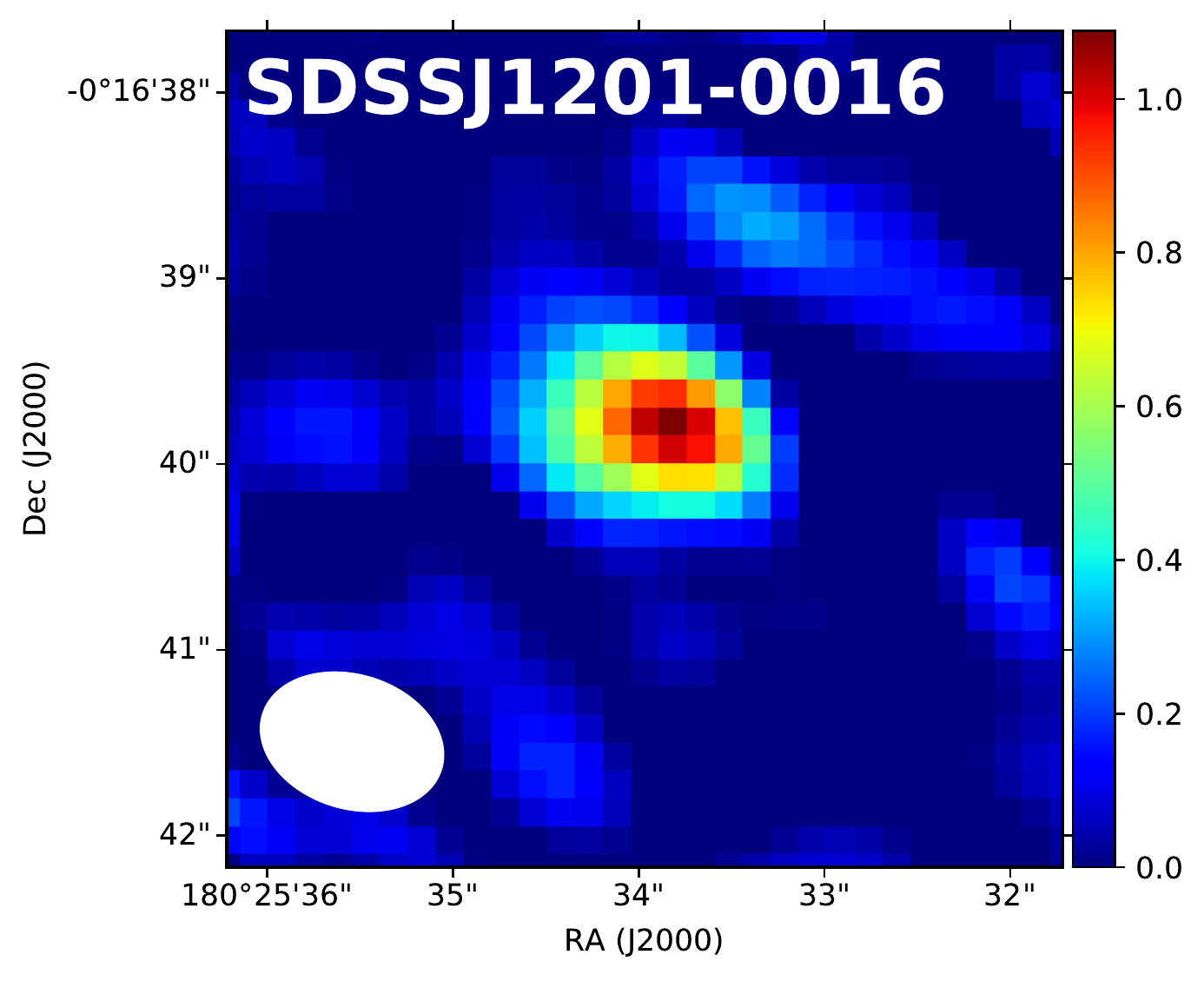}}
\put(90,148){\includegraphics[width=\wi]{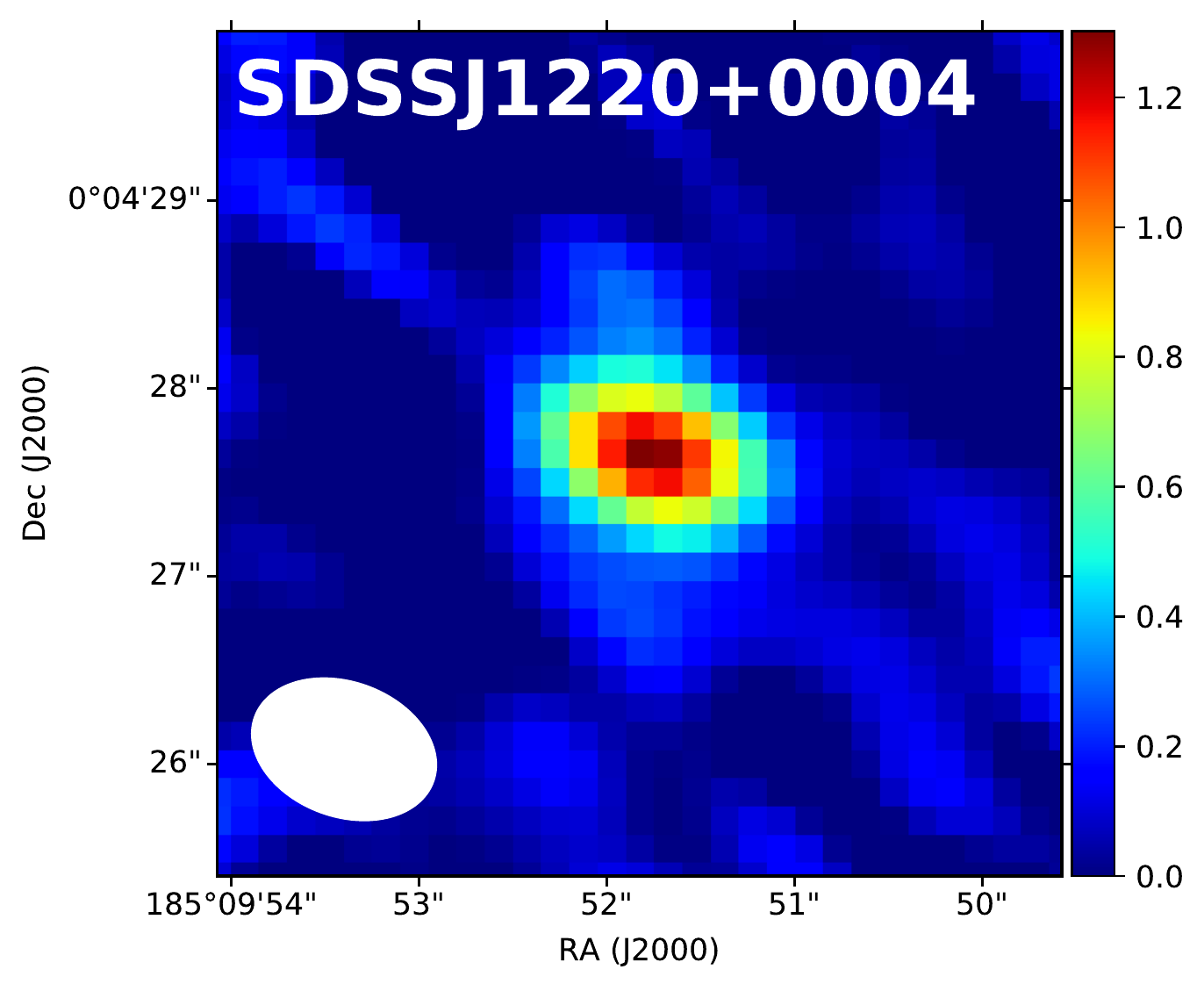}}
\put(135,148){\includegraphics[width=\wi]{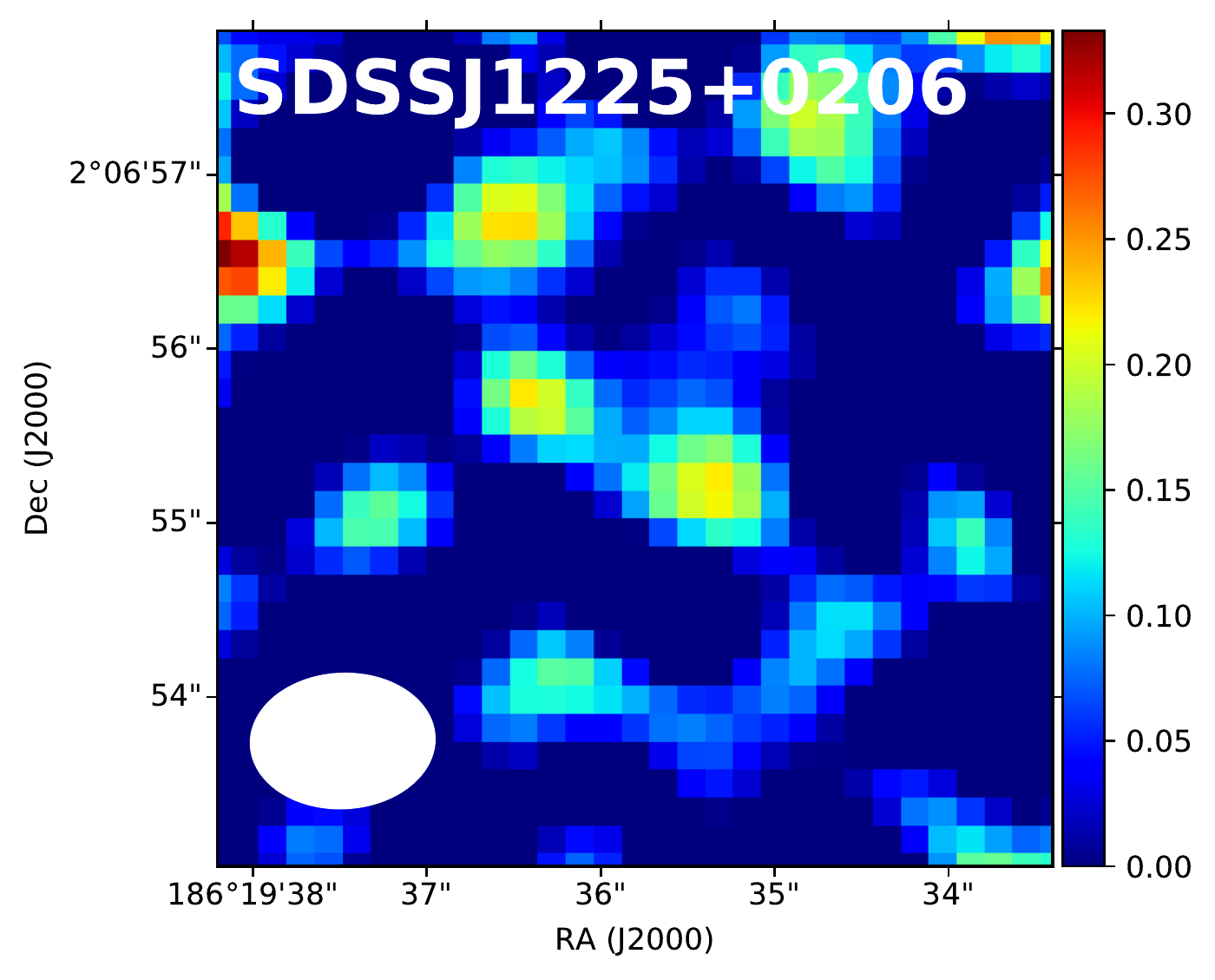}}

\put(0,111){\includegraphics[width=\wi]{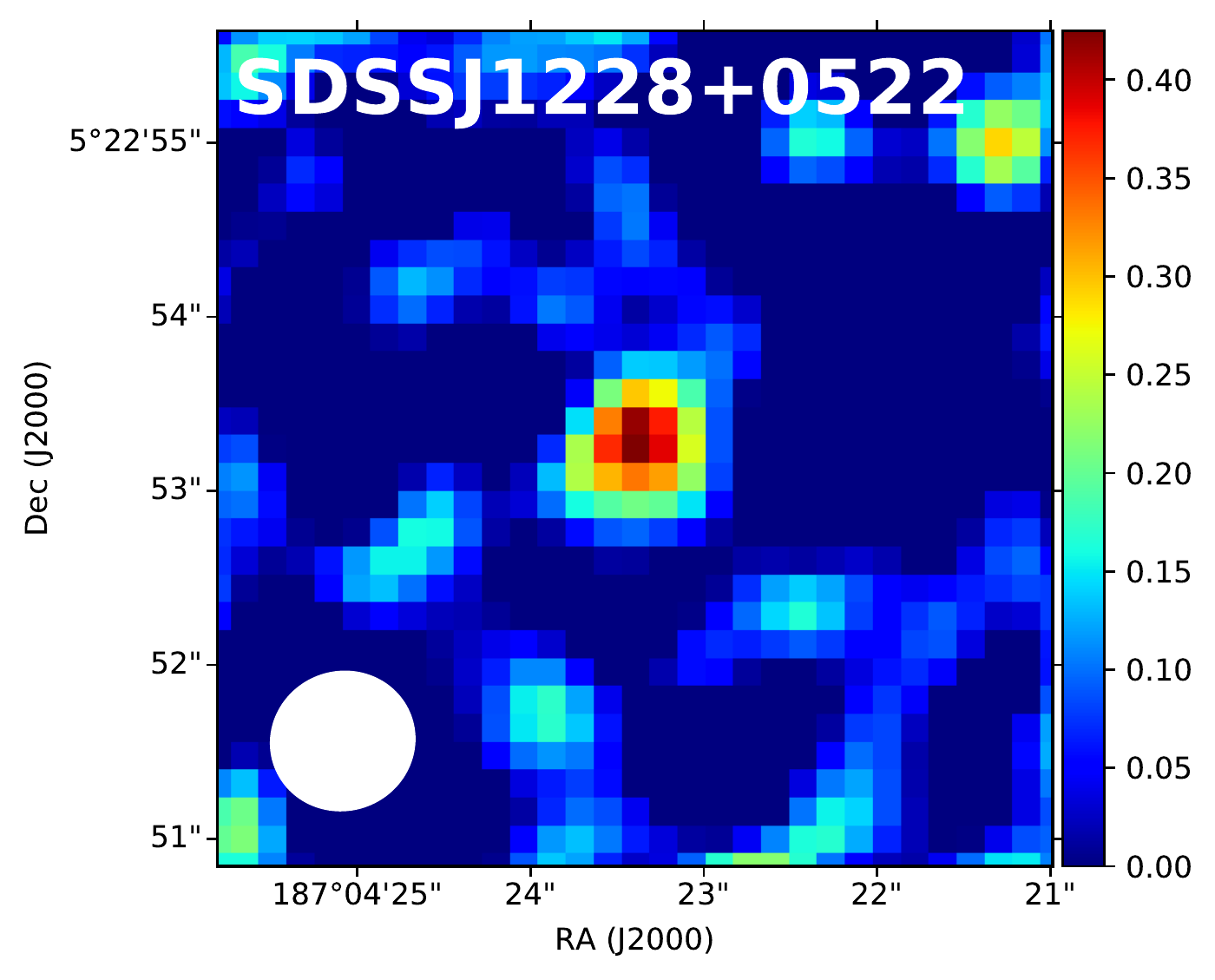}}
\put(45,111){\includegraphics[width=\wi]{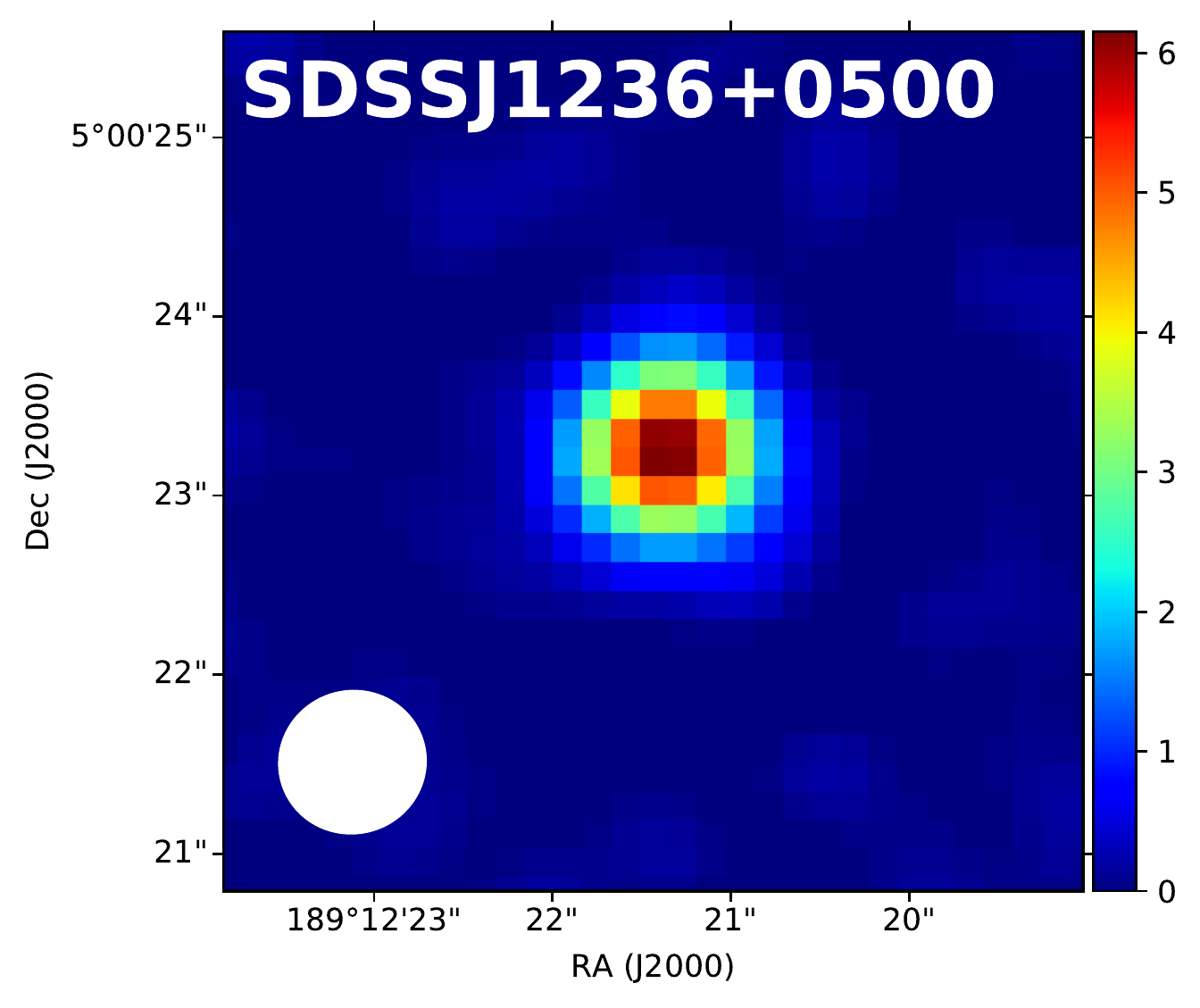}}
\put(90,111){\includegraphics[width=\wi]{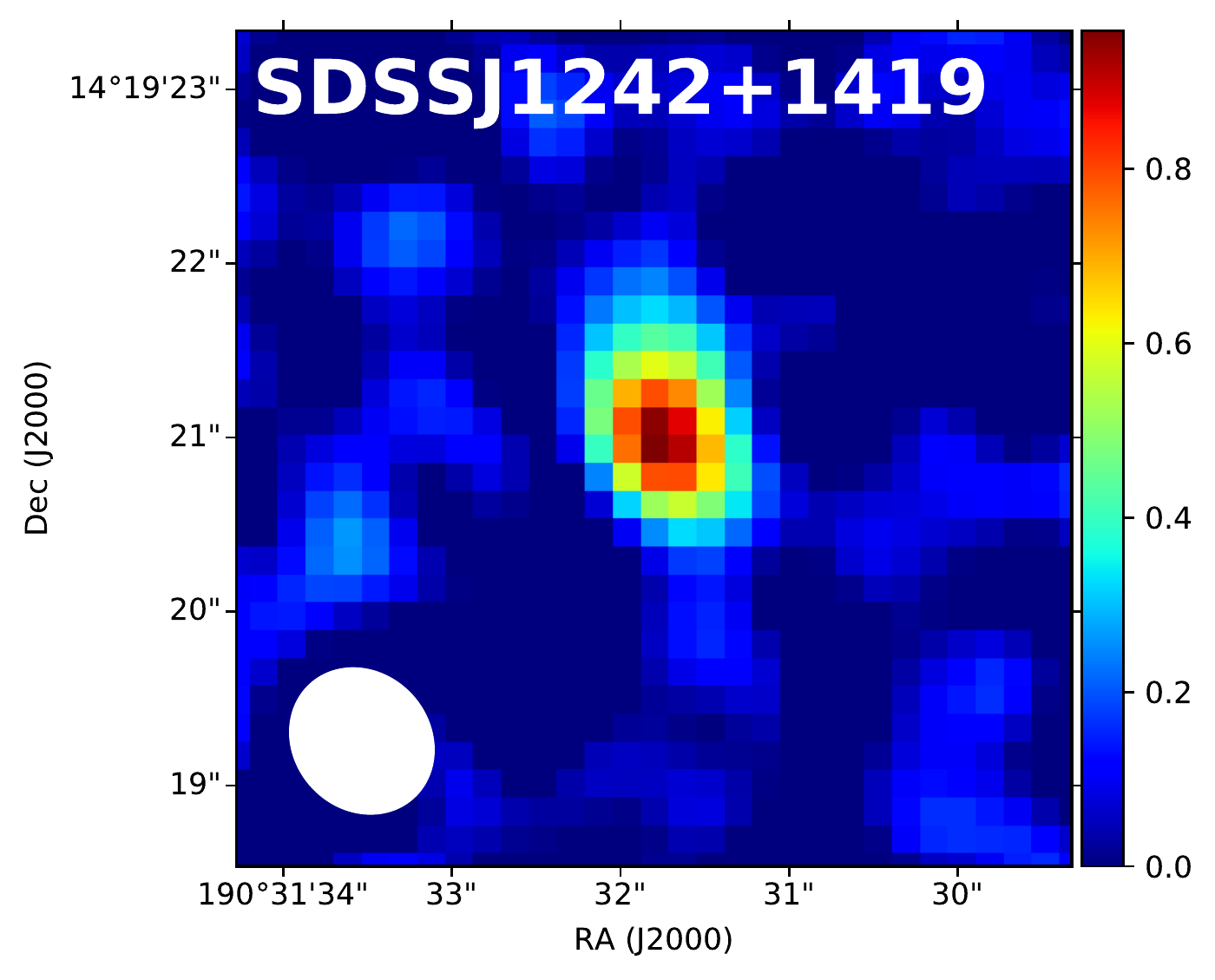}}
\put(135,111){\includegraphics[width=\wi]{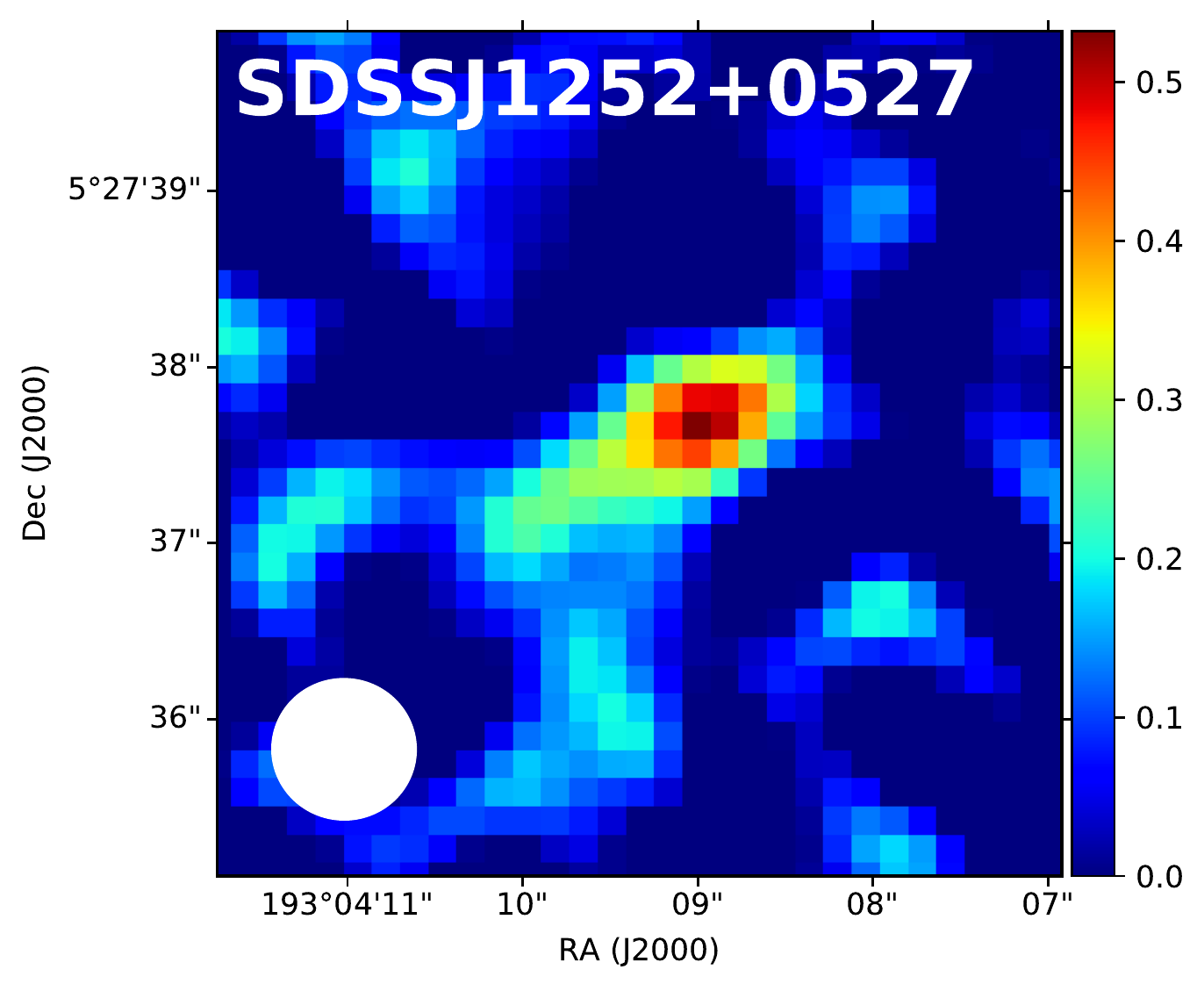}}

\put(0,74){\includegraphics[width=\wi]{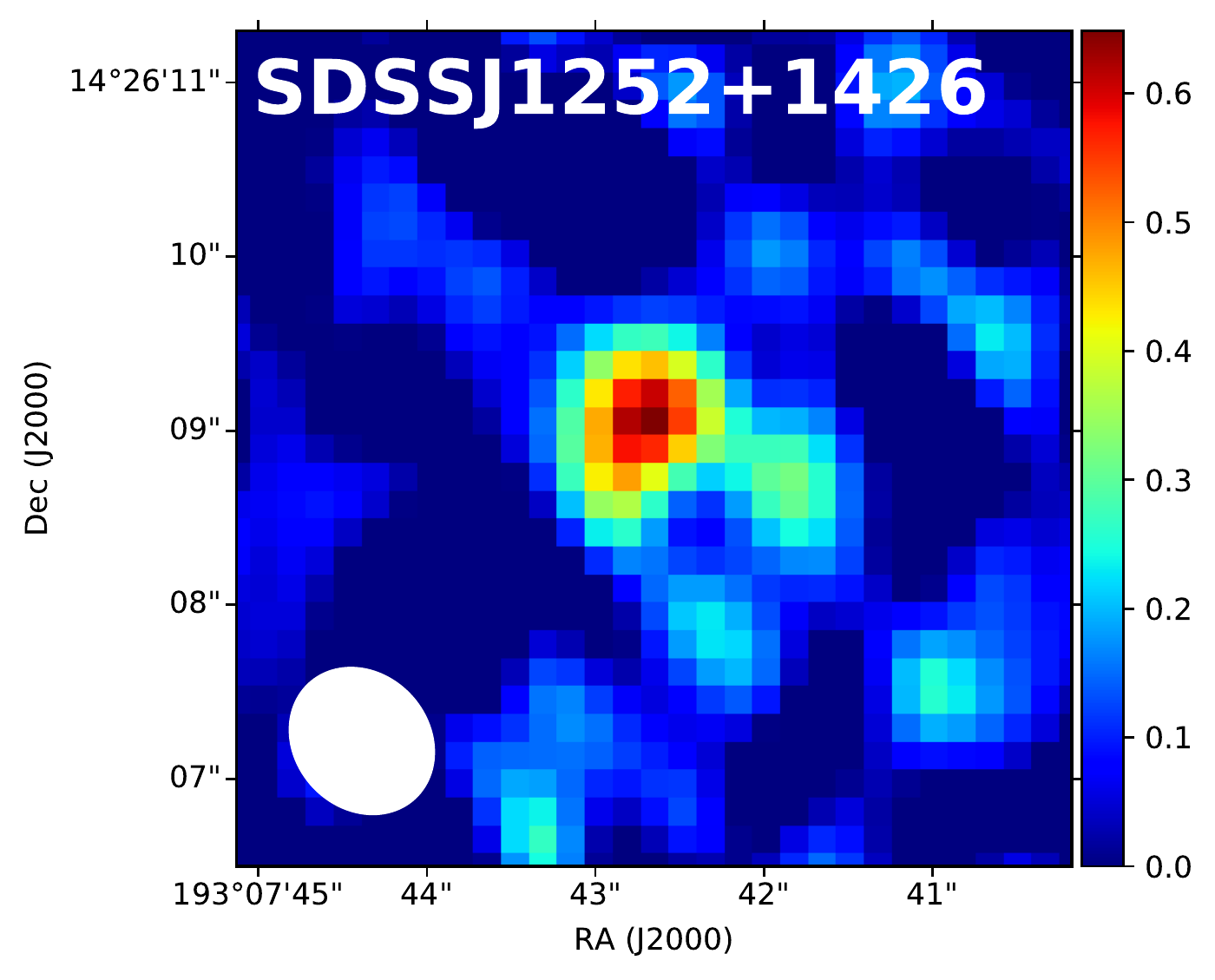}}
\put(45,74){\includegraphics[width=\wi]{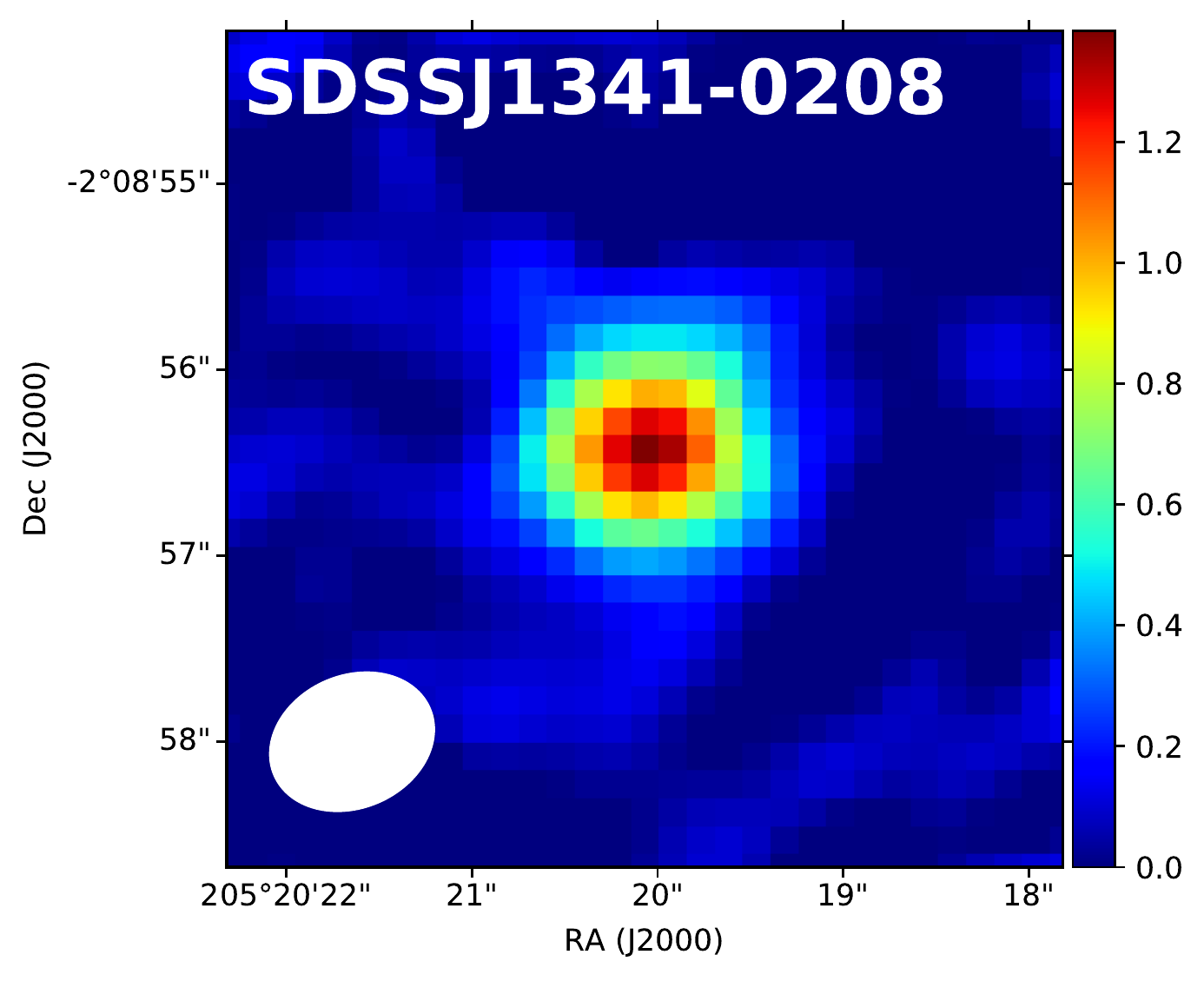}}
\put(90,74){\includegraphics[width=\wi]{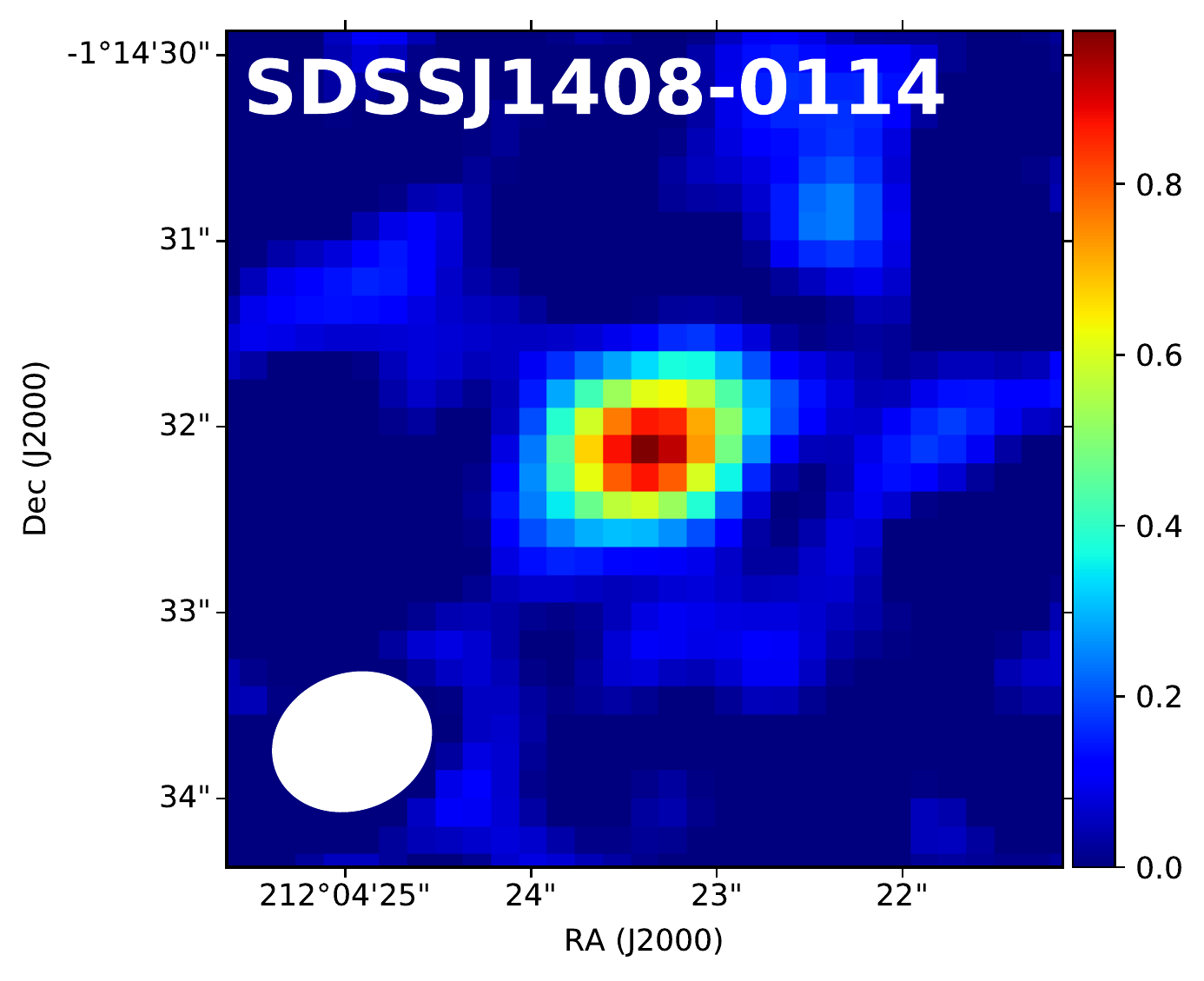}}
\put(135,74){\includegraphics[width=\wi]{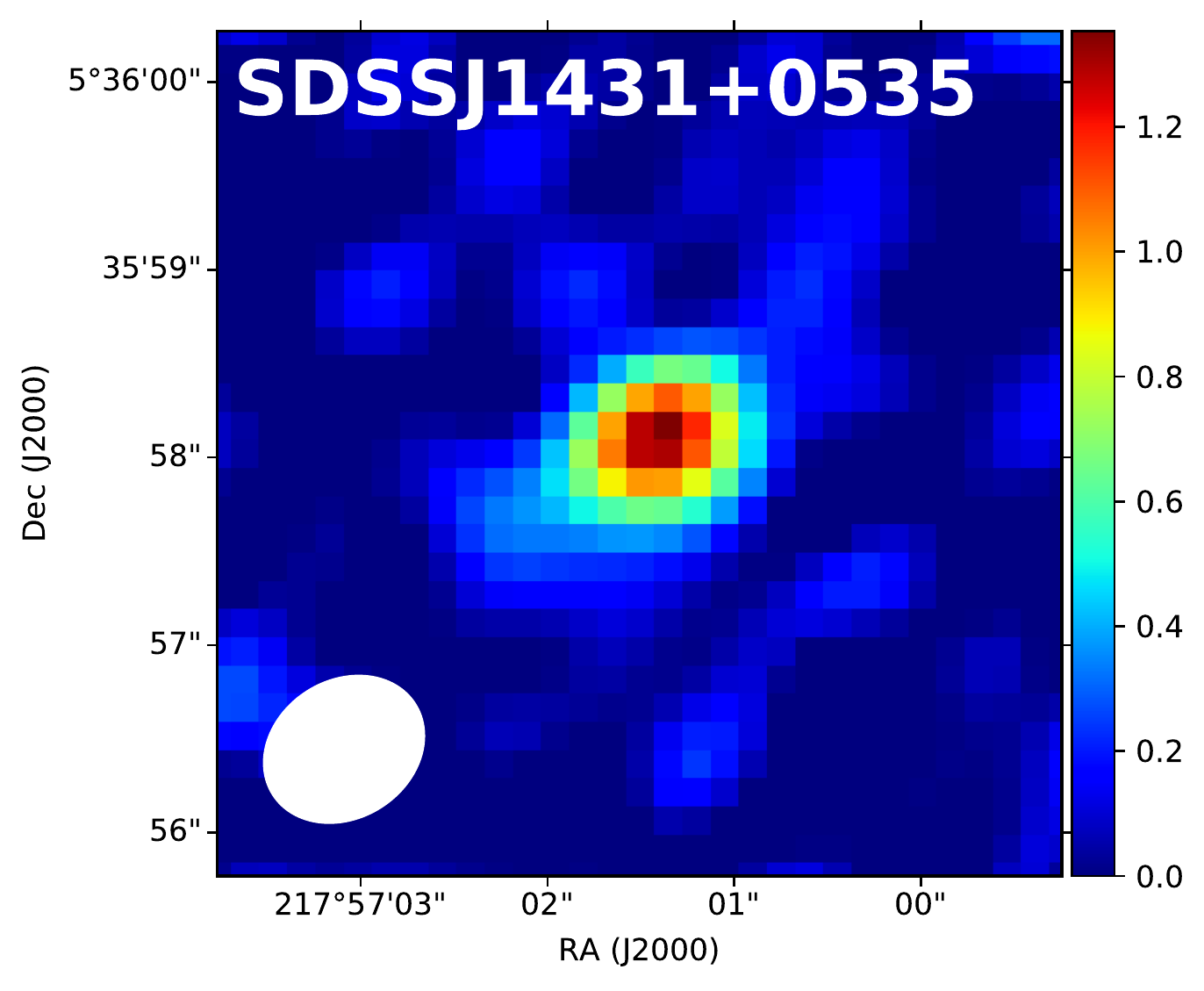}}

\put(0,37){\includegraphics[width=\wi]{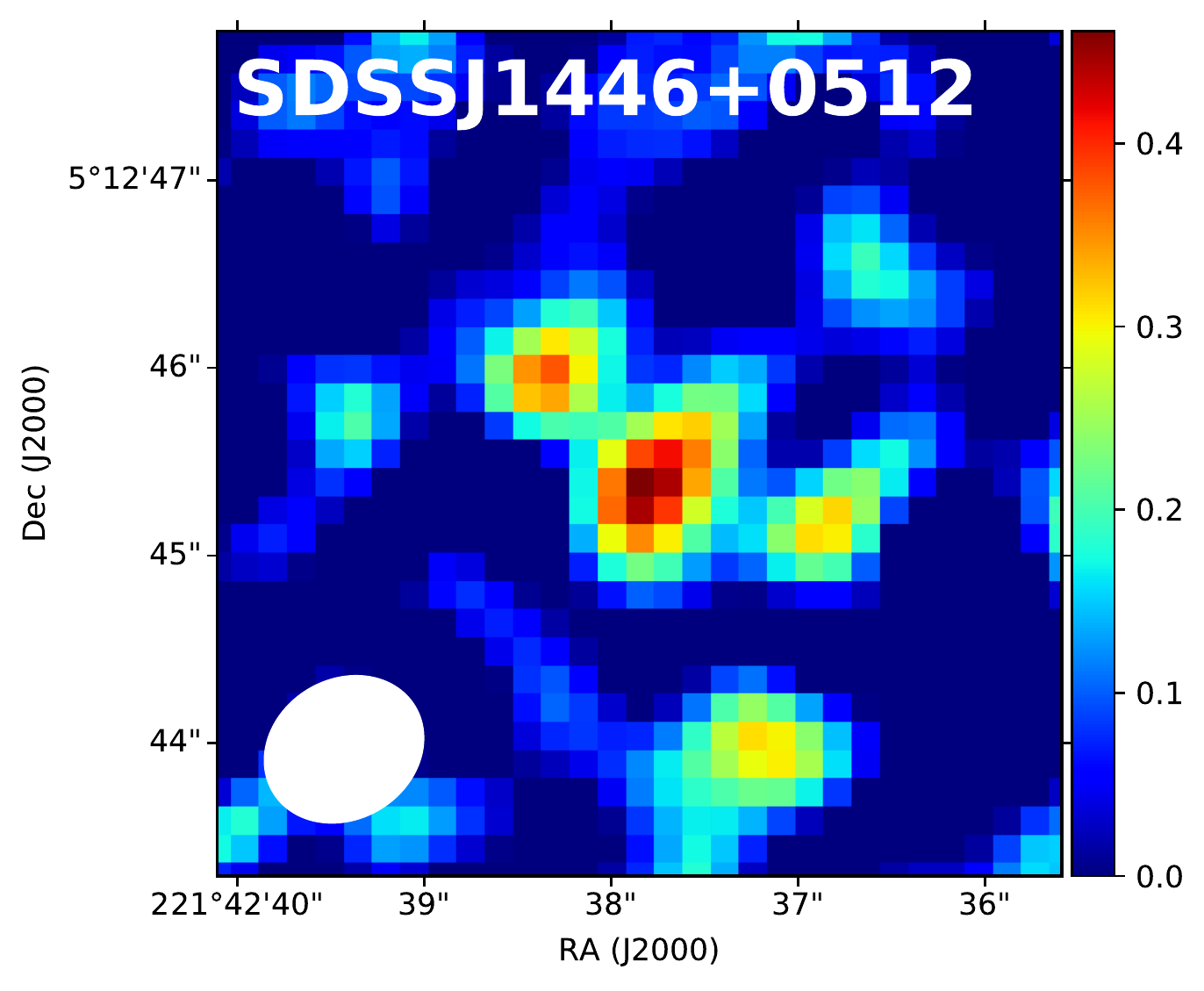}}
\put(45,37){\includegraphics[width=\wi]{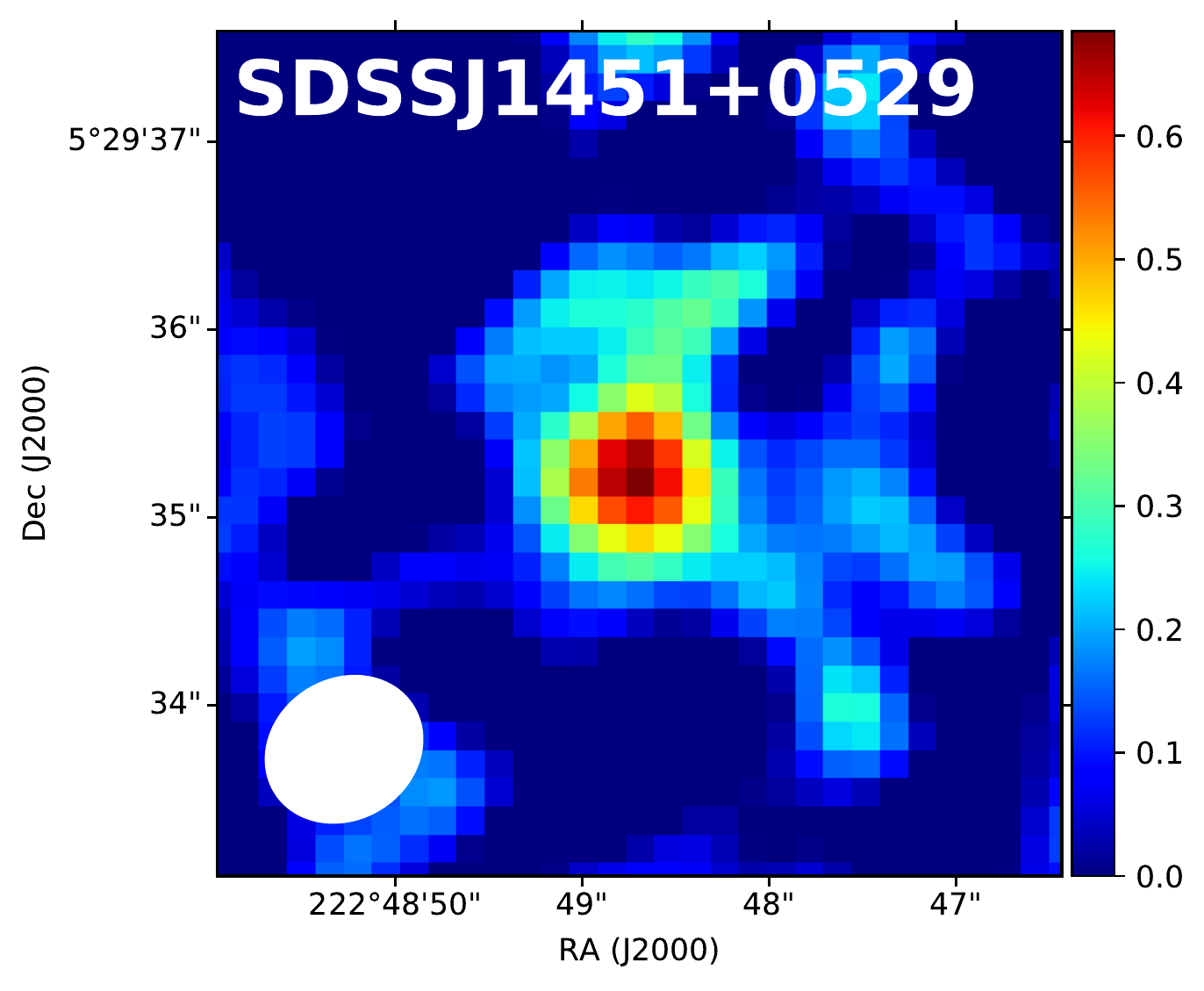}}
\put(90,37){\includegraphics[width=\wi]{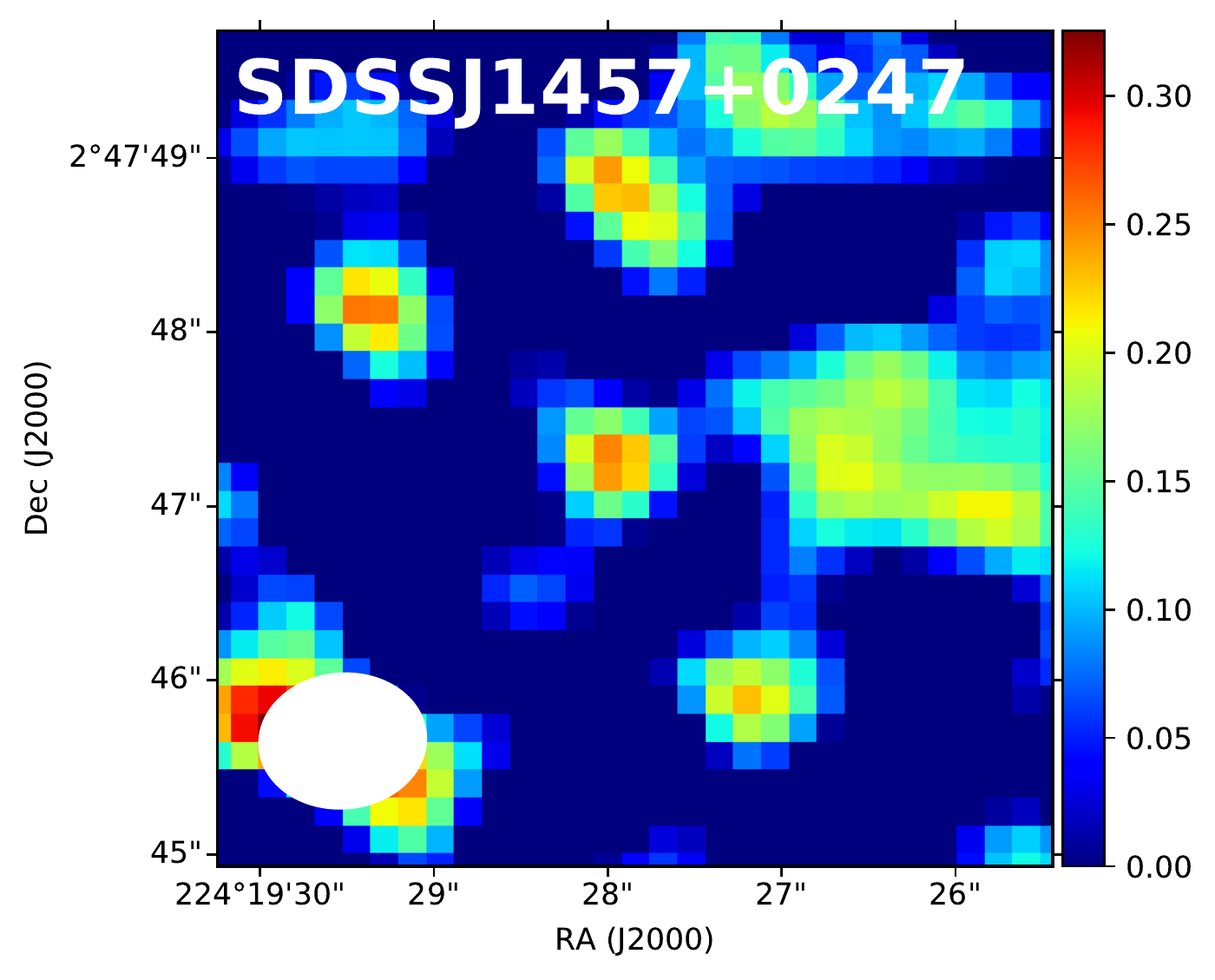}}
\put(135,37){\includegraphics[width=\wi]{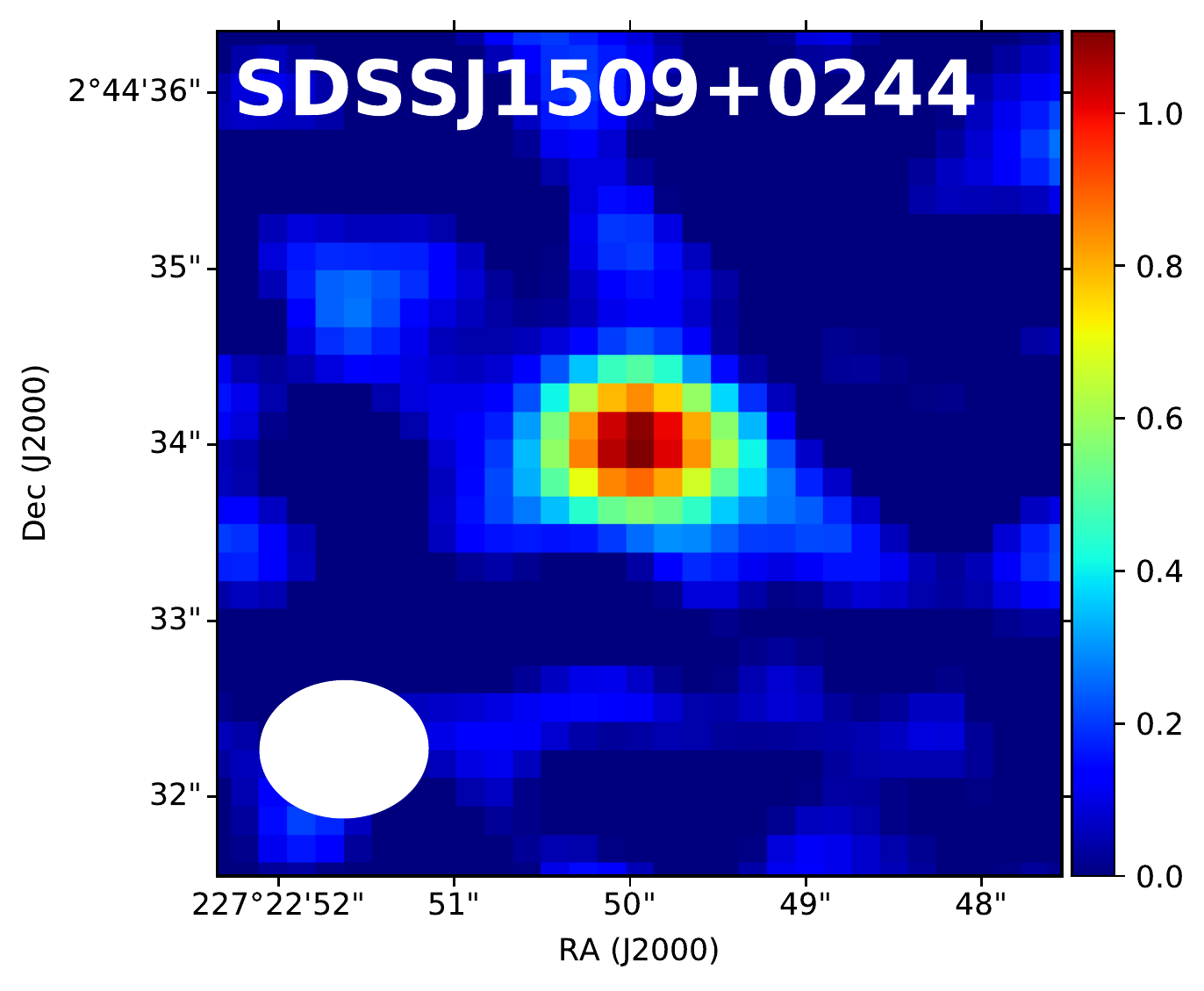}}

\put(0,0){\includegraphics[width=\wi]{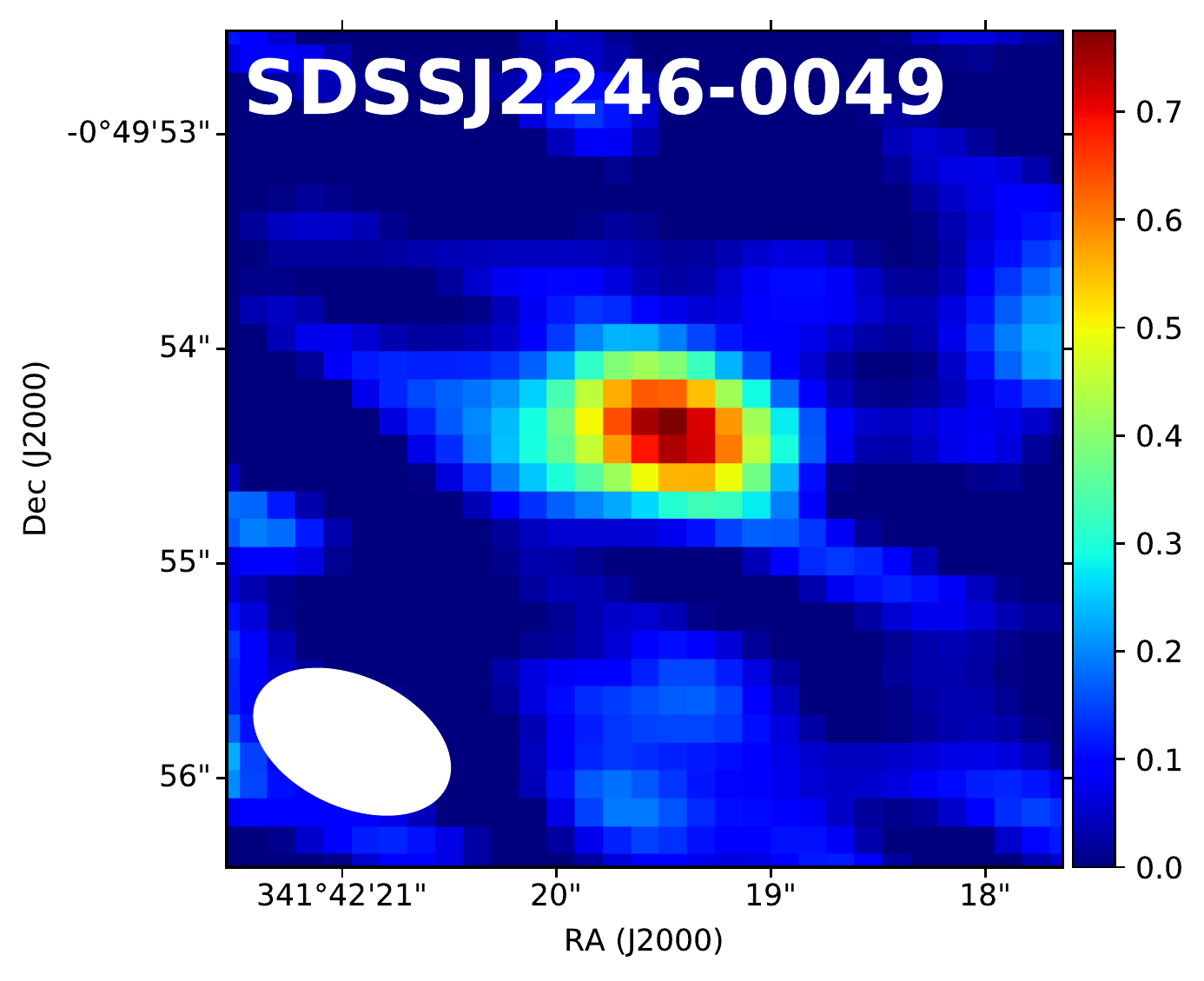}}
\put(45,0){\includegraphics[width=\wi]{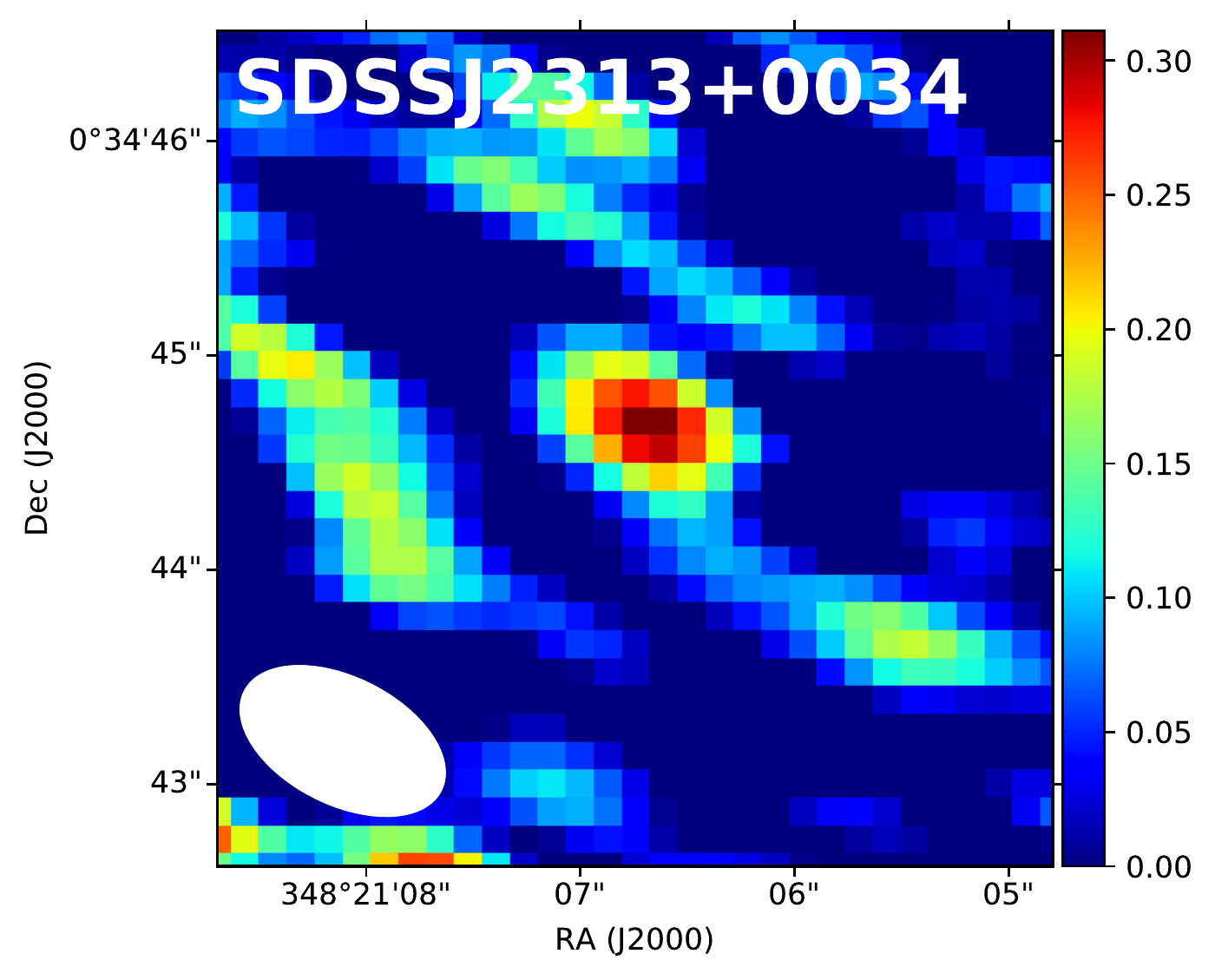}}
\put(90,0){\includegraphics[width=\wi]{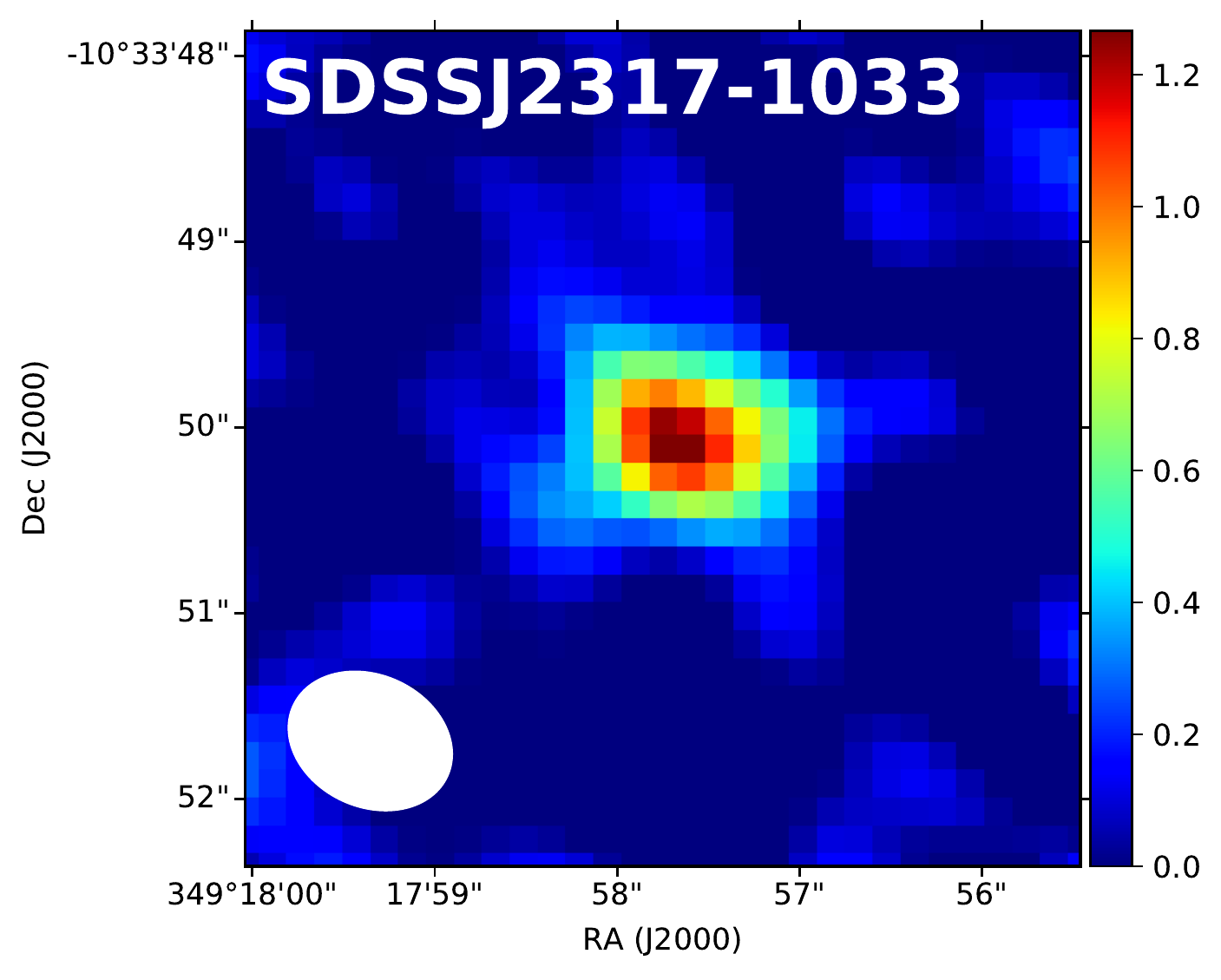}}
\put(135,0){\includegraphics[width=\wi]{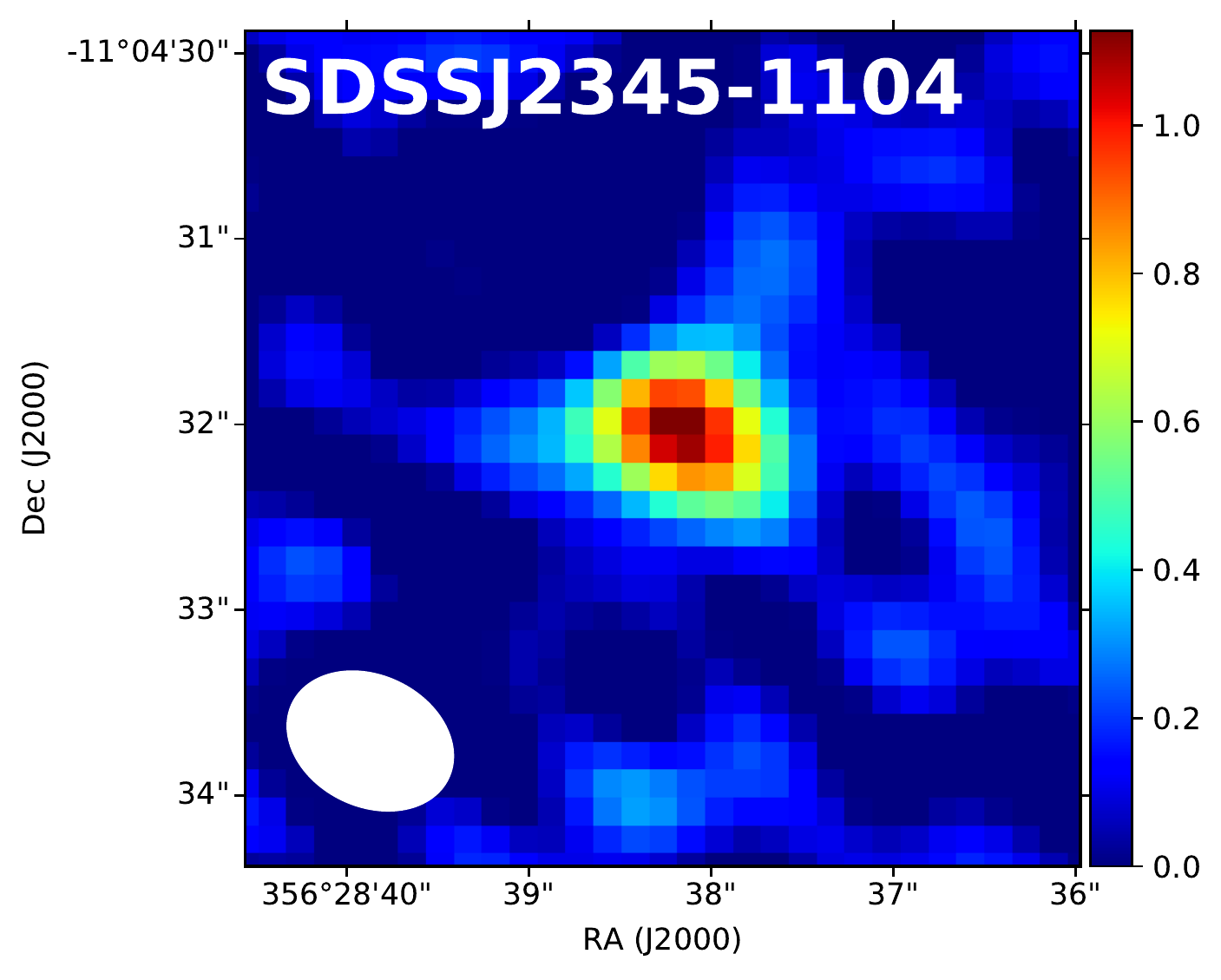}}
\end{picture}
\caption{ALMA Band 7 continuum images for the quasar sample. The color scale gives the flux in mJy. We show the beam size as white ellipse in the lower left corner. }
\label{fig:image_qso}
\end{figure*}

\rev{We show the Band 7 continuum images of all 20 QSO targets  for visualisation purposes in Fig.~\ref{fig:image_qso}. These images are based on the delivered data products using the ALMA pipeline (CASA version 5.1.1), applying a shallow clean.}
In total 19 of the 20 targets are detected at $850\mu$m at more than $3\sigma$ at the optical QSO position, 16 of them at more than $5\sigma$. The remaining source,  SDSS J1225+0206,  has a $1\sigma$ flux measurement of 0.1~mJy, just above our sensitivity limit and is considered a non-detection. While we are extracting sources blindly, we ensure that the flux is not boosted by noise for faint sources. For all detections their position is less than $\sim1/2$ of the beam away from the QSO position, typically $<0.2\arcsec$.
7 of the 20 quasars are spatially resolved, while the rest are un-resolved given the $\sim0.9\arcsec$ beam.

\rev{
For our sample, we found 4 additional sources within the ALMA beam\footnote{ One source each in the beam of SDSS J1225+0206, SDSS J1228+0522, SDSS J2246$-$0049 and SDSS J2313+0034}. None of them is detected in the optical SDSS images. A detailed discussion of these sources is beyond the scope of this paper and will be presented elsewhere. Here, we briefly discuss the implications on the multiplicity of AGN in sub-mm observations. 
The majority of our sample (80\%) have a unique source within the ALMA beam. Only 20\% have multiple sources. This is consistent with the recent results of \citet{Hatziminaoglou:2018} for FIR bright quasars. They found multiple sources in $\sim30$\% of their sample. This supports their suggestion that on average the majority of optically bright quasars is not triggered by early-stage mergers, but the results of other processes, and extends it the quasar population over a broader range of SFR. This suggestion is also consistent with quasar host galaxy studies at $z\sim2$ with HST \citep{Mechtley:2016,Marian:2019}. \citet{Trakhtenbrot:2017} reported on ALMA observations of 6 luminous quasars at $z\sim4.8$ and found three of them to have a companion source in the ALMA beam. Recent observations of a larger sample by this group lead to a multiplicity fraction around $\sim30\%$, more in line with our results \citep[see discussion in][]{Hatziminaoglou:2018}.
}

\subsection{Herschel data}
Out of our sample, 5 objects \rev{are covered by \textit{Herschel} maps} in the \textit{Herschel Science Archive}. Three of them have been observed as part of H-ATLAS \citep{Eales:2010}, and the other two fall into fields targeting nearby galaxies. Four of them have \textit{PACS} and \textit{SPIRE} data, while the remaining quasar only has \textit{SPIRE} coverage.
We retrieved the raw \textit{Herschel} data from the archive and reprocessed them following the method in \citet{Liu:2015}, but with the latest calibration in Herschel Interactive Processing Environment (HIPE) version 14. None of the targets are detected in either the \textit{PACS} or \textit{SPIRE} bands. The typical $3\sigma$ upper limits are $\sim40$~mJy beam$^{-1}$ at 350$\mu$m, corresponding to a luminosity of $\sim10^{46}$ erg s$^{-1}$ at $z\sim2$. Thus, as shown in section~\ref{sec:agnsed}, the \textit{Herschel} upper limits are not constraining for the SED of our objects. Therefore, we do not further consider these \textit{Herschel} upper limits in the following analysis.

\begin{table*}
\caption{Sample.}
\label{tab:sample}
\centering
\begin{tabular}{lcccccccccc}
\hline \hline 
Name &  R.A. & Decl. & $z$ & $i_{\rm{AB}}$ &  $M_i(z=2)$ &  FWHM(\ion{Mg}{ii})  & $\log L_{3000}$  & $\log \mbh$  & $\log \er$ & $\log L_\mathrm{bol}$ \\
&  (J2000) & (J2000) &  & (mag) &  (mag) &  (km s$^{-1}$)  & (erg s$^{-1}$)  &($M_\odot$)  &  & (erg s$^{-1}$) \\
(1) & (2) & (3) & (4) & (5) & (6) & (7) & (8) & (9) & (10) & (11) \\
\hline 
SDSS J1149+0151 & 11:49:00.344 & +01:51:17.07 &    2.066 &  17.41 & $-$28.38 & $3196\pm161$ &  46.56  &   9.34 &  $-$0.38 &  47.07 \\
SDSS J1201$-$0016 & 12:01:42.253 & $-$00:16:39.85 &    1.993 &  17.93 & $-$27.67 & $5073\pm2241$ &  46.46  &   9.67 &  $-$0.82 &  46.96 \\
SDSS J1220+0004 & 12:20:39.452 & +00:04:27.67 &    2.048 &  17.15 & $-$28.63 & $4083\pm194$ &  46.67  &   9.62 &  $-$0.55 &  47.17 \\
SDSS J1225+0206 & 12:25:18.409 & +02:06:56.61 &    2.026 &  17.03 & $-$28.73 & $3548\pm174$ &  46.64  &   9.48 &  $-$0.44 &  47.15 \\
SDSS J1228+0522 & 12:28:17.556 & +05:22:53.52 &    2.021 &  17.48 & $-$28.26 & $2811\pm142$ &  46.66  &   9.29 &  $-$0.23 &  47.16 \\
SDSS J1236+0500 & 12:36:49.431 & +05:00:23.31 &    1.941 &  17.54 & $-$28.06 & $3266\pm156$ &  46.51  &   9.33 &  $-$0.42 &  47.02 \\
SDSS J1242+1419 & 12:42:06.112 & +14:19:21.04 &    1.973 &  16.96 & $-$28.68 & $3716\pm255$ &  46.60  &   9.49 &  $-$0.50 &  47.10 \\
SDSS J1252+0527 & 12:52:16.586 & +05:27:37.77 &    1.903 &  16.96 & $-$28.59 & $3548\pm103$ &  46.80  &   9.58 &  $-$0.38 &  47.31 \\
SDSS J1252+1426 & 12:52:30.846 & +14:26:09.22 &    1.938 &  16.48 & $-$29.13 & $4784\pm474$ &  46.91  &   9.90 &  $-$0.60 &  47.41 \\
SDSS J1341$-$0208 & 13:41:21.339 & $-$02:08:56.43 &    2.095 &  17.02 & $-$28.80 & $3696\pm565$ &  46.76  &   9.59 &  $-$0.43 &  47.27 \\
SDSS J1408$-$0114 & 14:08:17.560 & $-$01:14:32.11 &    1.945 &  17.44 & $-$28.18 & $3801\pm148$ &  46.48  &   9.44 &  $-$0.56 &  46.99 \\
SDSS J1431+0535 & 14:31:48.094 & +05:35:58.09 &    2.095 &  16.46 & $-$29.39 & $5525\pm1237$ &  47.01  &  10.09 &  $-$0.69 &  47.51 \\
SDSS J1446+0512 & 14:46:50.519 & +05:12:45.45 &    2.082 &  17.57 & $-$28.25 & $3611\pm1063$ &  46.47  &   9.39 &  $-$0.52 &  46.98 \\
SDSS J1451+0529 & 14:51:15.245 & +05:29:35.19 &    2.053 &  17.15 & $-$28.63 & $2973\pm177$ &  46.72  &   9.37 &  $-$0.26 &  47.22 \\
SDSS J1457+0247 & 14:57:17.856 & +02:47:47.44 &    1.976 &  16.26 & $-$29.41 & $2715\pm104$ &  46.84  &   9.37 &  $-$0.13 &  47.34 \\
SDSS J1509+0244 & 15:09:31.329 & +02:44:34.00 &    2.047 &  17.70 & $-$28.10 & $3700\pm589$ &  46.48  &   9.41 &  $-$0.54 &  46.98 \\
SDSS J2246$-$0049 & 22:46:49.299 & $-$00:49:54.37 &    2.038 &  17.29 & $-$28.62 & $4308\pm156$ &  46.70  &   9.68 &  $-$0.59 &  47.20 \\
SDSS J2313+0034 & 23:13:24.456 & +00:34:44.52 &    2.087 &  15.90 & $-$30.01 & $3260\pm138$ &  47.17  &   9.73 &  $-$0.17 &  47.67 \\
SDSS J2317$-$1033 & 23:17:11.843 & $-$10:33:50.11 &    2.004 &  17.37 & $-$28.30 & $4041\pm845$ &  46.56  &   9.54 &  $-$0.58 &  47.07 \\
SDSS J2345$-$1104 & 23:45:54.543 & $-$11:04:32.06 &    1.948 &  17.67 & $-$27.92 & $4124\pm277$ &  46.41  &   9.47 &  $-$0.66 &  46.92 \\
\hline
\end{tabular}
\flushleft
(1) Shortened SDSS name; (2) right ascension; (3) declination; (4) Redshift taken from \citet{Hewett:2010}; (5) $i$-band AB magnitude from SDSS; (6) absolute magnitude in $i$ at $z=2$ taken from \citet{Shen:2011}; 
\end{table*}

\subsection{Optical spectral properties, black hole masses and bolometric luminosities} \label{sec:mbh}
Optical spectra, covering the \ion{Mg}{II} and \ion{C}{IV} broad line region, are available for all 20 objects from SDSS DR7 \citep{Abazajian:2009}. In addition, BOSS spectra are available for 9 quasars in SDSS DR14 \citep{Abolfathi:2018}, which provide a wider wavelength coverage and an improved quality especially around \ion{Mg}{II}. \rev{Thereby, the use of the new BOSS spectra improves the spectral measurements of} \ion{Mg}{II} \rev{and $L_{3000}$, which are essential for the black hole mass estimate and bolometric luminosity. Furthermore, we are interested in shape measurements (e.g., asymmetry) of the \ion{C}{IV} line profile, which are not provided in the SDSS DR7 black hole mass catalog \citep{Shen:2011}. This made it necessary to perform a dedicated, consistent  and visually verified spectral modeling of the latest available spectra for this sample and to re-calculate the black hole masses and bolometric luminosities in a consistent manner. For this, we use the primary spectra given in the SDSS DR14 quasar catalog \citep{Paris:2018}. We use the improved redshifts from \citet{Hewett:2010} for our study.}

We perform spectral model fits independently to the \ion{Mg}{II} and \ion{C}{IV} line regions, including a power law continuum, iron emission template and a multi-Gaussian model for the broad emission lines. Details on the line fitting are given in Appendix~\ref{sec:specfit}, and we show the optical spectra and the best-fit models in Figures~\ref{fig:spec_m2} and \ref{fig:spec_c4}. We measure the full width at half maximum (FWHM) of the broad \ion{Mg}{II}  line from the multi-Gaussian fit and the continuum luminosity at 3000\AA{}\ $L_{3000}$ from the power law continuum and report these in Table~\ref{tab:sample}.

Black hole mass estimates can be obtained using the virial method \citep[e.g][]{McLure:2002,Vestergaard:2006}. Under the assumption of virialized motion of the broad line region (BLR) gas and using established empirical scaling relations between continuum luminosity and BLR size \citep[e.g.][]{Kaspi:2005,Bentz:2009}, this method allows an estimate of the mass of the SMBH, with a typical uncertainty of $\sim$0.3~dex. While the broad H$\beta$ line is generally considered as the most reliable black hole mass estimator, the \ion{Mg}{ii} line has also been shown to provide robust black hole masses \citep{McGill:2008,Trakhtenbrot:2012,Mejia:2016,Woo:2018,Schulze:2018b}. Since H$\beta$ is not available for our sources, we base our black hole mass estimates on \ion{Mg}{ii}. Specifically, we use the relation from \citet{Shen:2011}

\begin{equation}
\mbh (\rm{MgII})= 10^{6.74} \left( \frac{L_{3000}}{10^{44}\,\mathrm{erg\,s}^{-1}}\right)^{0.62} \left( \frac{\mathrm{FWHM}}{1000\,\mathrm{km\,s}^{-1} }\right)^2 M_\odot
\end{equation} 

While we also observe and fit the \ion{C}{IV} line, this line is known to be a poor SMBH mass estimator, especially for very luminous AGN \citep{Baskin:2005,Shen:2012b,Trakhtenbrot:2012}, due to a non-virial component, potentially associated with an outflow, thus we do not use this line as a black hole mass estimator. We rather use it as a tracer of AGN winds, indicated by (1) the blue shift between the centroid of the \ion{Mg}{ii} line, which is known to be close to the systemic redshift \citep{Shen:2016}, and \ion{C}{IV}; (2) the line asymmetry of the \ion{C}{IV} line. We further see the presence of intrinsic absorption features associated with the \ion{C}{IV} line in a subset of our targets.

Our measure of the bolometric luminosity is based on applying a bolometric correction factor $f_{\rm bol}$ to the continuum luminosity $L_{3000}$, i.e. $L_{\rm bol}=f_{\rm bol} L_{3000}$. A commonly adopted value is a constant factor $f_{\rm bol}=5.15$ \citep{Richards:2006}. However, this value most likely overestimates $L_{\rm bol}$, especially for luminous quasars \citep{Marconi:2004, Trakhtenbrot:2012}. Our estimate does include emission from the dust torus, which represents reprocessed emission and thus should be excluded. Furthermore, the AGN SED is luminosity dependent, which should lead to a decrease in $f_{\rm bol, 3000}$ with increasing UV-luminosity. Therefore, we use the luminosity-dependent $f_{\rm bol}$ prescription for $L_{3000}$ by \citet[][their equation~5]{Trakhtenbrot:2012}, based on the bolometric corrections by \citet{Marconi:2004}, which considers these effects. At $L_{3000}>10^{46}$~erg s$^{-1}$  $f_{\rm bol}$ flattens at a value of $\sim3.2$. Since all quasars in our sample have $L_{3000}$ above this value we adopt a constant bolometric correction factor $f_{\rm bol}=3.2$  for our sample. Note that our bolometric luminosities are systematically lower by a factor 1.6 compared to the use of the commonly adopted value of $f_{\rm bol}=5.15$.
The Eddington ratio is given by $\er=L_{\rm{bol}}/L_{\rm{Edd}}$, where $L_{\rm{Edd}}\cong1.3\times 10^{38} (\mbh / M_\odot)$~erg s$^{-1}$ is the Eddington luminosity for the object, given its black hole mass.

We show the location of our sample in the SMBH mass-luminosity plane in Figure~\ref{fig:sample} in relation to the general SDSS DR7 quasar population at the same redshift. With this selection, our sample constitutes the most luminous sources in the SDSS DR7 quasar catalog, with $\log L_{\rm bol}>46.9$ [erg s$^{-1}$], black hole masses $\log \mbh>9.2$ [$M_\odot$] and Eddington ratios above 10\%.

\begin{figure*}
\centering
\includegraphics[width=17cm,clip]{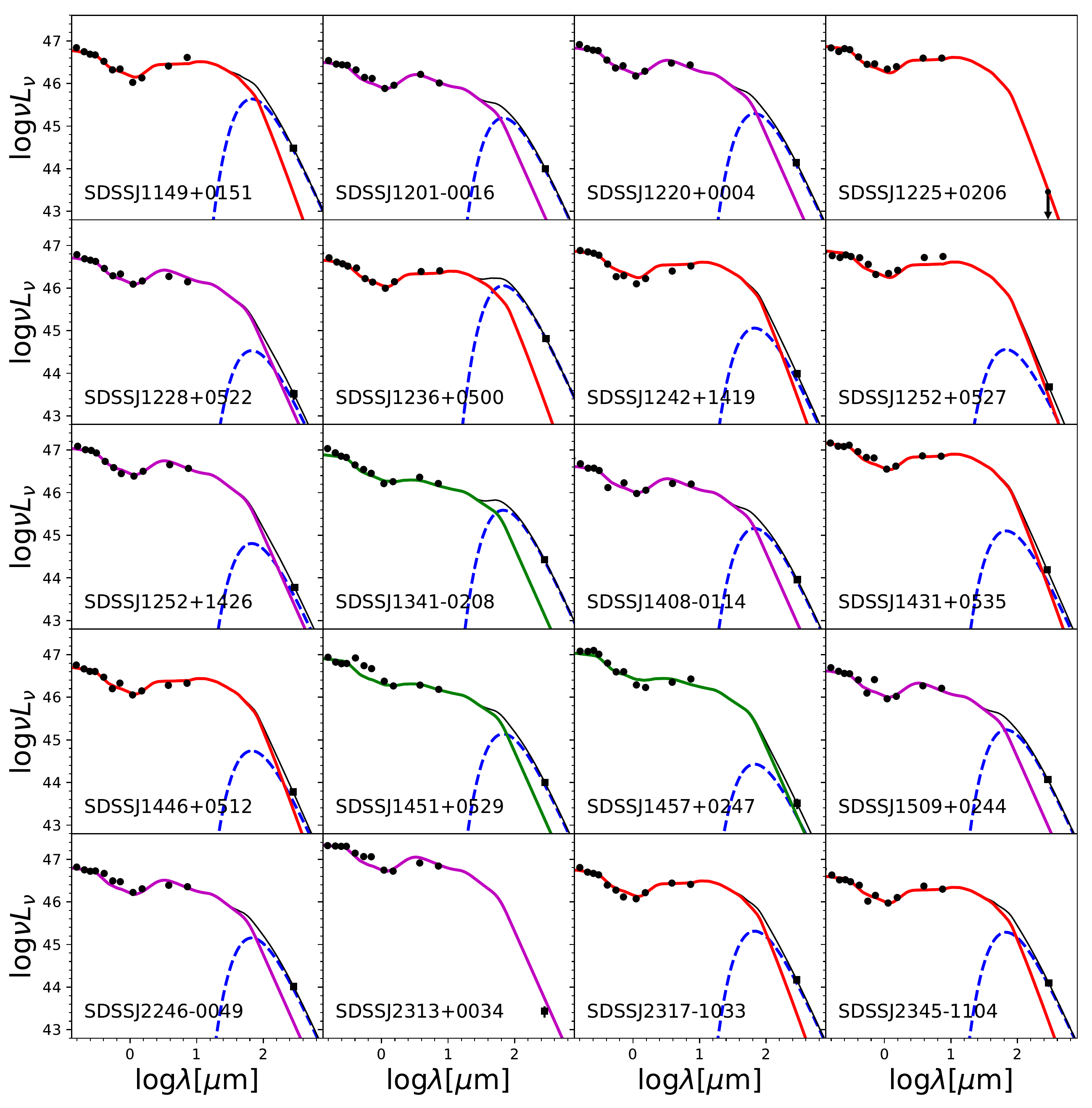} 
\caption{Spectral energy distribution (SED) and SED decomposition results for our luminous quasar sample. The black circles show the optical to mid-IR photometry from SDSS, 2MASS and WISE. The ALMA data point at $850\mu$m is shown as a black square. The solid lines give the best fit AGN SED template to the optical to mid-IR photometry, based on the three templates provided in \citet{Lyu:2017}: 1) a normal quasar (red line), 2) a WDD quasar (magenta line) and 3) a HDD quasar (green line). The best-fit modified black body to the AGN-subtracted ALMA flux is shown by the blue dashed line. The thin black line represents the total (AGN+SF) SED.
}
\label{fig:sed} 
\end{figure*}

\begin{figure*}
\centering
\includegraphics[width=8cm,clip]{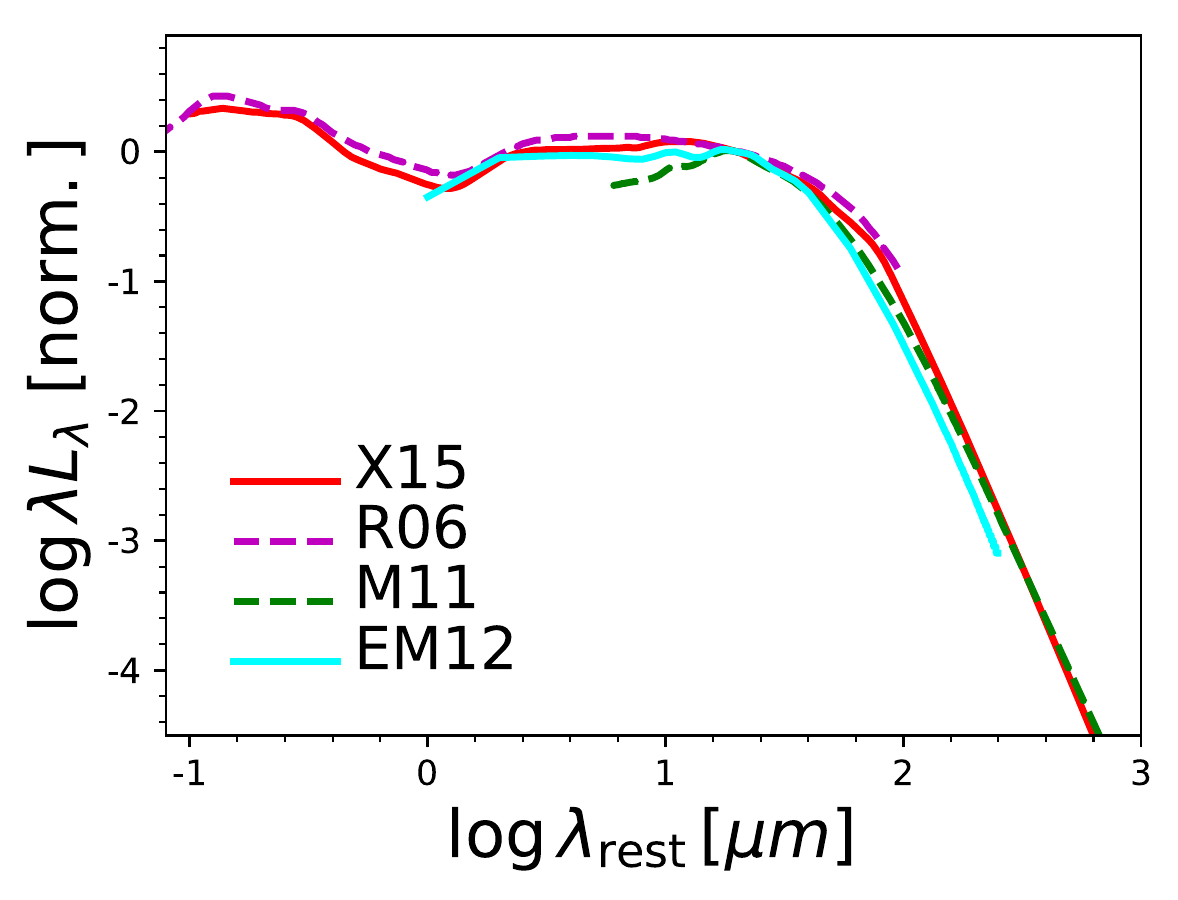} \hspace{0.5cm}
\includegraphics[width=8cm,clip]{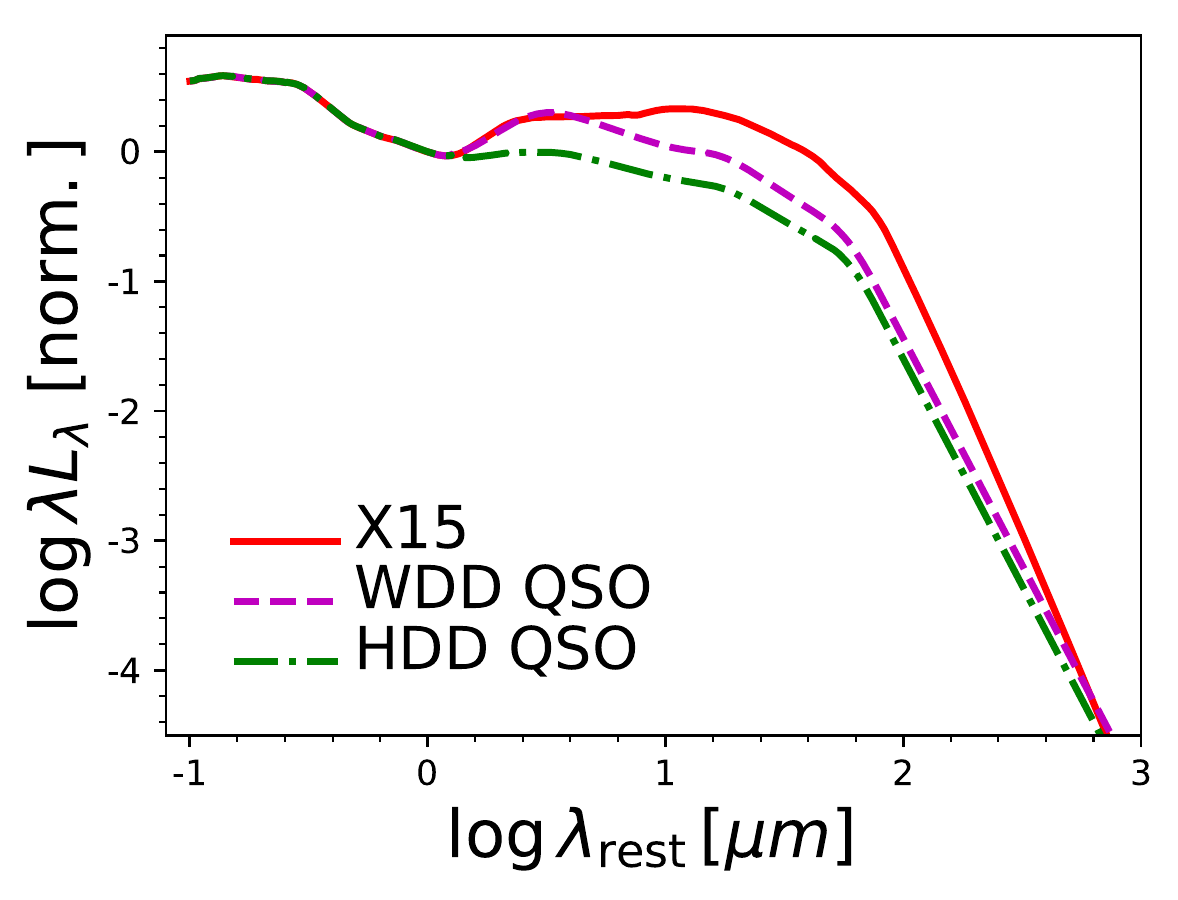} 
\caption{Left panel: Examples of empirical AGN SED templates proposed in the literature, normalized at 20$\mu$m. These are the template by \citet[][red solid line]{Xu:2015}, \citet[][magenta dashed line]{Richards:2006}, \citet[][green dashed line]{Mullaney:2011},
and the extended version of the SED by \citet{Mor:2012}, as presented in \citet[][EM12, cyan solid line]{Lani:2017}. Right panel: Comparison of the standard AGN SED template by \citet[][red solid line]{Xu:2015} to the AGN SEDs templates of warm-dust-deficient (WDD, magenta dashed line) quasars and hot-dust-deficient (HDD, green dotted-dashed line) quasars, as presented by \citet{Lyu:2017b}. All templates are normalized at $1\mu$m.
}
\label{fig:sedtemplates} 
\end{figure*}

\section{Analysis} \label{sec:results}

\subsection{AGN spectral energy distribution} \label{sec:agnsed}
Due to the high luminosity of our sample, the quasar emission will dominate the total luminosity at almost all wavelengths. This is particularly true for the UV, optical, mid-IR, and even around $60\mu$m in the FIR. However, at $850\mu$m, the quasar emission is minimized and most likely sub-dominant or even fully negligible. Thus, we investigate the SEDs of our sample to assess the potential AGN contribution at $850\mu$m. 

We make use of multi-wavelength photometry, as provided in the SDSS DR14 quasar catalog \citep{Paris:2018}. This includes optical photometry in $ugriz$ from SDSS \citep{Albareti:2017}, $JHK$ near-IR photometry from the Two Micron All Sky Survey \citep[2MASS,][]{Skrutskie:2006} and mid-IR data from the Wide-Field Infrared Survey \citep[WISE,][]{Wright:2010} at $3.4, 4.6, 12$ and $22 \mu$m. We omit the $u$-band data, since at the redshift of our objects it is severely contaminated by Ly$\alpha$ emission and Ly$\alpha$ forest absorption. We do not consider the \textit{Herschel} upper limits for the fit. We note that they are always significantly above our best-fit SED. We also do not include the ALMA data in the fit. We show the photometric data points for our sample in Figure~\ref{fig:sed}. 

Several different AGN SED templates have been used in previous studies to decompose the AGN and the host galaxy emission. In Figure~\ref{fig:sedtemplates} we show a few commonly adopted AGN SED templates \citep{Richards:2006,Mor:2012,Mullaney:2011,Xu:2015}, each normalized at 20$\mu$m. The SED template by \citet{Richards:2006} covers the UV-mid-IR range and does not fully extend into the FIR. 
The AGN templates by \citet{Mullaney:2011} (their high luminosity template), \citet{Mor:2012} (as provided in \citealt{Lani:2017}) and \citet{Xu:2015}(as provided in \citealt{Lyu:2017}) are largely consistent at $\lambda_{\rm rest}>20\mu$m.\footnote{Contrary, the SED template proposed by \citet{Symeonidis:2016} would predict a significant AGN contribution from cold dust at $\lambda_{\rm rest}>100\mu$m. However, their results have been challenged more recently \citep{Lyu:2017,Lani:2017,Stanley:2018}. Our ALMA observations of extremely luminous quasars seem to support the concerns raised by these authors. In 18/20 cases our measured 850$\mu$m flux is below the expectation from the template by \citet{Symeonidis:2016}, when the latter is anchored to the quasar photometry at $\lambda_{\rm rest}=60\mu$m. Thus, we do not further consider their SED template in our study.}
In the following, we use the AGN template by \citet{Xu:2015}, since it provides the widest wavelength coverage from the UV to the FIR. This enables us to use the full UV to mid-IR photometry available to constrain the AGN SED. The \citet{Xu:2015} template is based on the original AGN SED by \citet{Elvis:1994}, removing the star formation contribution from that SED. In the UV to mid-IR regime it is consistent with the more recent quasar SED by \citet{Richards:2006}.

While the SED for the majority of luminous quasars is well represented by this SED template, \citet{Lyu:2017b} demonstrated that there are significant sub-populations ($30-40$\%) whose SED deviates from the standard SED, primarily in the mid-IR due to a dust-deficiency.  \citet{Lyu:2017b} characterize them as 1) hot-dust-deficient (HDD) quasars ($15-23$\%), showing very weak emission from the NIR all the way to the FIR, and 2) 
warm-dust-deficient (WDD) quasars ($14-17$\%), with similar  NIR emission as normal quasars, but relatively weak emission in the mid-IR to FIR \citep[see also][]{Hao:2010,Hao:2011,Mor:2011,Jun:2013}. We show their respective SED templates in the right panel of Figure~\ref{fig:sedtemplates}. 

To characterize the AGN SED for our sample, we use a simple approach by fitting with a library of SEDs consisting of the three AGN SED temples provided in \citet{Lyu:2017}: 1) the standard AGN template by \citet{Xu:2015}, 2) a warm-dust-deficient (WDD) quasar template and 3) a hot-dust-deficient (HDD) quasar template. We consider these templates to broadly cover the diversity of expected UV to mid-IR SED shapes for our sample. For each template, the absolute normalization is the only free parameter in our $\chi^2$ minimization routine. We then choose the SED template with the smallest $\chi^2$ value as the best fit for each object in our sample. \rev{In all cases the difference in $\chi^2$ is significant ($\Delta \chi^2_{\mathrm{reduced}}>>1$).}
We do not consider the host galaxy contribution to the UV to mid-IR emission, since for our very luminous quasars it is clearly sub-dominant over the full wavelength range.

The best fit results for the AGN SED are shown in Figure~\ref{fig:sed}. From the 20 luminous quasars in our sample, 9 are fitted with a standard SED, 8 with a WDD SED and 3 with a HDD SED. The latter corresponds to a fraction of 15\%, consistent with previous work \citep{Hao:2011,Mor:2011,Lyu:2017}. The fraction of 40\% modeled as WDD in our sample is higher than reported for low-$z$ PG quasars \citep{Lyu:2017}. However, they also find a strong increase of the WDD fraction with luminosity, generally consistent with our results. This is also consistent with the observation of an anti-correlation between the mid-IR to optical luminosity ratio with luminosity \citep{Maiolino:2007,Treister:2008,Roseboom:2013,Duras:2017}. 
In the context of this paper, we mainly use the best fit AGN SED to estimate and subtract the AGN contribution to the ALMA 850$\mu$m continuum flux. 

\begin{table*}
\caption{ALMA measurements}
\label{tab:res}
\centering
\begin{tabular}{lcccccccc}
\hline \hline 
Name &   $S_{850\mu\mathrm{m}}$  & $\log L_{850\mu\mathrm{m},{\rm Tot}}$ & $\log L_{850\mu\mathrm{m},{\rm CD}}$ & $\log L_\mathrm{IR,MBB}$  & ${SFR}_\mathrm{MBB}$ & $\log L_\mathrm{IR,temp}$  & ${SFR}_\mathrm{temp}$ & Size \\
&  (mJy) & (erg s$^{-1}$) & (erg s$^{-1}$) & (erg s$^{-1}$)& ($M_\odot$ yr$^{-1}$) & (erg s$^{-1}$) & ($M_\odot$ yr$^{-1}$) & ($\arcsec$)  \\
(1) & (2) & (3) & (4) & (5) & (6) & (7) & (8) & (9)\\
\hline 
SDSS J1149+0151 &  $  2.79\pm  0.11$  & 44.48 &  44.44 & $46.33^{+ 0.11}_{- 0.73}$ & $  553^{+  154}_{-  450}$  & $46.06^{+ 0.10}_{- 0.14}$ & $  301^{+   80}_{-   83}$ & $0.29\pm0.07$ \\ \noalign{\smallskip}
SDSS J1201$-$0016 &  $  0.99\pm  0.09$  & 44.00 &  43.96 & $45.89^{+ 0.11}_{- 0.74}$ & $  203^{+   57}_{-  166}$  & $45.63^{+ 0.10}_{- 0.14}$ & $  110^{+   30}_{-   31}$  & $-$\\ \noalign{\smallskip}
SDSS J1220+0004 &  $  1.31\pm  0.09$  & 44.14 &  44.09 & $45.99^{+ 0.11}_{- 0.73}$ & $  251^{+   70}_{-  205}$  & $45.72^{+ 0.10}_{- 0.14}$ & $  137^{+   37}_{-   38}$   & $-$\\ \noalign{\smallskip}
SDSS J1225+0206 &  $<  0.28$  &  $<43.46$ & $<43.46$ & $<45.37$ & $<   61$ & $<45.10$ & $<   33$  & $-$ \\  \noalign{\smallskip}
SDSS J1228+0522 &  $  0.32\pm  0.09$  & 43.52 &  43.32 & $45.23^{+ 0.11}_{- 0.74}$ & $   44^{+   12}_{-   36}$  & $44.97^{+ 0.10}_{- 0.14}$ & $   24^{+    6}_{-    7}$  & $-$\\ \noalign{\smallskip}
SDSS J1236+0500 &  $  7.04\pm  0.09$  & 44.82 &  44.81 & $46.77^{+ 0.11}_{- 0.74}$ & $ 1514^{+  426}_{- 1239}$  & $46.50^{+ 0.10}_{- 0.14}$ & $  815^{+  221}_{-  231}$ & $0.26\pm0.02$ \\ \noalign{\smallskip}
SDSS J1242+1419 &  $  1.01\pm  0.10$  & 43.99 &  43.82 & $45.77^{+ 0.11}_{- 0.74}$ & $  151^{+   43}_{-  124}$  & $45.50^{+ 0.10}_{- 0.14}$ & $   82^{+   22}_{-   23}$  & $-$\\ \noalign{\smallskip}
SDSS J1252+0527 &  $  0.54\pm  0.09$  & 43.68 &  43.29 & $45.27^{+ 0.11}_{- 0.74}$ & $   49^{+   14}_{-   40}$  & $45.00^{+ 0.10}_{- 0.15}$ & $   26^{+    7}_{-    8}$  & $-$\\ \noalign{\smallskip}
SDSS J1252+1426 &  $  0.64\pm  0.10$  & 43.77 &  43.56 & $45.52^{+ 0.11}_{- 0.74}$ & $   86^{+   24}_{-   70}$  & $45.25^{+ 0.10}_{- 0.15}$ & $   46^{+   13}_{-   13}$  & $-$\\ \noalign{\smallskip}
SDSS J1341$-$0208 &  $  2.38\pm  0.10$  & 44.43 &  44.40 & $46.28^{+ 0.11}_{- 0.73}$ & $  491^{+  137}_{-  400}$  & $46.01^{+ 0.10}_{- 0.14}$ & $  267^{+   71}_{-   73}$  & $0.71\pm0.06$\\ \noalign{\smallskip}
SDSS J1408$-$0114 &  $  0.97\pm  0.07$  & 43.96 &  43.91 & $45.87^{+ 0.11}_{- 0.74}$ & $  193^{+   54}_{-  158}$  & $45.60^{+ 0.10}_{- 0.14}$ & $  104^{+   28}_{-   29}$  & $-$\\ \noalign{\smallskip}
SDSS J1431+0535 &  $  1.39\pm  0.09$  & 44.19 &  43.92 & $45.79^{+ 0.11}_{- 0.73}$ & $  161^{+   45}_{-  131}$  & $45.53^{+ 0.10}_{- 0.14}$ & $   87^{+   23}_{-   24}$  & $-$\\ \noalign{\smallskip}
SDSS J1446+0512 &  $  0.55\pm  0.09$  & 43.78 &  43.55 & $45.43^{+ 0.11}_{- 0.73}$ & $   70^{+   20}_{-   57}$  & $45.17^{+ 0.10}_{- 0.14}$ & $   38^{+   10}_{-   11}$  & $-$\\ \noalign{\smallskip}
SDSS J1451+0529 &  $  0.94\pm  0.12$  & 44.00 &  43.94 & $45.83^{+ 0.11}_{- 0.73}$ & $  177^{+   50}_{-  144}$  & $45.57^{+ 0.10}_{- 0.14}$ & $   96^{+   26}_{-   27}$  & $0.51\pm0.17$\\ \noalign{\smallskip}
SDSS J1457+0247 &  $  0.34\pm  0.09$  & 43.51 &  43.20 & $45.14^{+ 0.11}_{- 0.74}$ & $   36^{+   10}_{-   29}$  & $44.87^{+ 0.10}_{- 0.14}$ & $   19^{+    5}_{-    5}$  & $-$\\ \noalign{\smallskip}
SDSS J1509+0244 &  $  1.11\pm  0.09$  & 44.07 &  44.03 & $45.93^{+ 0.11}_{- 0.73}$ & $  221^{+   62}_{-  181}$  & $45.67^{+ 0.10}_{- 0.14}$ & $  121^{+   32}_{-   33}$  & $-$\\ \noalign{\smallskip}
SDSS J2246$-$0049 &  $  0.98\pm  0.08$  & 44.01 &  43.95 & $45.85^{+ 0.11}_{- 0.73}$ & $  184^{+   51}_{-  150}$  & $45.59^{+ 0.10}_{- 0.14}$ & $  100^{+   27}_{-   28}$  & $0.33\pm0.11$\\ \noalign{\smallskip}
SDSS J2313+0034 &  $  0.25\pm  0.07$  & $43.44$ &  $<43.44$ & $<45.31$ & $<   54$ & $<45.05$ & $<   29$   & $-$\\  \noalign{\smallskip}
SDSS J2317$-$1033 &  $  1.47\pm  0.10$  & 44.17 &  44.09 & $46.01^{+ 0.11}_{- 0.74}$ & $  268^{+   75}_{-  219}$  & $45.75^{+ 0.10}_{- 0.14}$ & $  146^{+   39}_{-   41}$  & $0.26\pm0.11$\\ \noalign{\smallskip}
SDSS J2345$-$1104 &  $  1.34\pm  0.11$  & 44.10 &  44.04 & $46.00^{+ 0.11}_{- 0.74}$ & $  257^{+   72}_{-  210}$  & $45.73^{+ 0.10}_{- 0.14}$ & $  139^{+   38}_{-   39}$  & $0.40\pm0.11$\\ \noalign{\smallskip}
\hline
\end{tabular}
\flushleft
(1) SDSS target name; (2) measured ALMA flux at $850\mu$m or 3$\sigma$ upper limit; (3) corresponding $850\mu$m luminosity; (4) $850\mu$m luminosity after subtraction of the AGN contribution; (5)-(6) total IR luminosity ($8-1000\mu$m) and corresponding SFR assuming a modified blackbody for the cold dust emission; (7)-(8) total IR luminosity and corresponding SFR using a SF-galaxy template based on the library of \citet{Magdis:2012}. (9) Source FWHM for the resolved sources in the sample, based on our used circular Gaussian model.
\end{table*}

\subsection{FIR luminosities and Star formation rates}
In this section, we use our ALMA continuum measurements to estimate the total FIR luminosity from star formation (SF; integrated over $8-1000\mu$m), hereafter \lir, and the SFR in the quasar host galaxies. Our measured ALMA Band-7 continuum fluxes and the corresponding luminosity are listed in Table~\ref{tab:res}. We subtract the AGN contribution from the AGN SED best-fit from this luminosity and use this AGN-subtracted luminosity throughout to probe the emission due to SF. 
We verified that our qualitative conclusions do not change if we would use the continuum fluxes without subtracting the AGN contribution instead. For SDSS~J2313+0034, the most luminous quasar in the sample, the expected AGN contribution is higher than the ALMA measurement, and we set an upper limit to $L_{850\mu{\rm m}}$ due to star formation at the ALMA measurement. For the rest of the sample we find an AGN contribution \rev{to $L_{850\mu\mathrm{m}}$ in the range of $3-59\%$, with a median value of 14\%. The AGN contribution shows a positive correlation with the quasar luminosity, with a Spearman rank order coefficient of $r_S=0.53$.
We also verified that our results do not depend on the specific choice of AGN template used to account for the AGN contribution. For our three adopted AGN templates we find a mean difference between them of 0.08~dex in $\log L_{850\mu\mathrm{m},{\rm CD}}$ or SFR.
}

Estimates of total \lir\, and SFR from a single measurement on the Rayleigh-Jeans-tail of the cold dust emission bears reasonably high uncertainties, due to the unknown dust temperature and dust mass. 
We use two independent approaches to estimate \lir\, to alleviate such uncertainties.
First, we assume a modified blackbody (MBB) spectrum to represent the thermal dust emission, presumably heated by SF:
\begin{equation}
L_{\rm mbb}(\nu) = N_{\rm mbb} \frac{ \nu^{3+\beta} }{e^{h \nu / k_b T_{\rm d}}-1}  
\end{equation}
We fix the emissivity index to $\beta=1.6$ and the dust temperature to $T_{\rm d}=47$~K  \citep{Beelen:2006} and determine the normalization $N_{\rm mbb}$ from the ALMA continuum flux measurement. These values are widely adopted in previous studies of high-$z$ quasar host galaxies \citep[e.g.][]{Wang:2013,Willott:2015,Izumi:2018,Decarli:2018}. \citet{Trakhtenbrot:2017} found that these assumptions provide results in good agreement with {\it Herschel} observations for a representative sample of luminous type-1 quasars at $z\sim4.8$  studied with ALMA. This adopted value for $\beta$ is also consistent with studies of local luminous and ultraluminous infrared galaxies \citep{Casey:2012}.
Other AGN host studies with detections at several wavelengths, in particular close to the peak of the blackbody emission using e.g. {\it Herschel}, are able to fit for $T_{\rm d}$ and typically report temperatures around $30-50$~K \citep{Beelen:2006,Wang:2008,Leipski:2013,Petric:2015,Ma:2015,Duras:2017}. We use this temperature range to derive lower and upper limits on \lir. We consider this uncertainty, introduced by the unknown dust temperature, to be the dominating source of uncertainty on \lir.

\begin{figure*}
\centering
\includegraphics[width=18cm,clip]{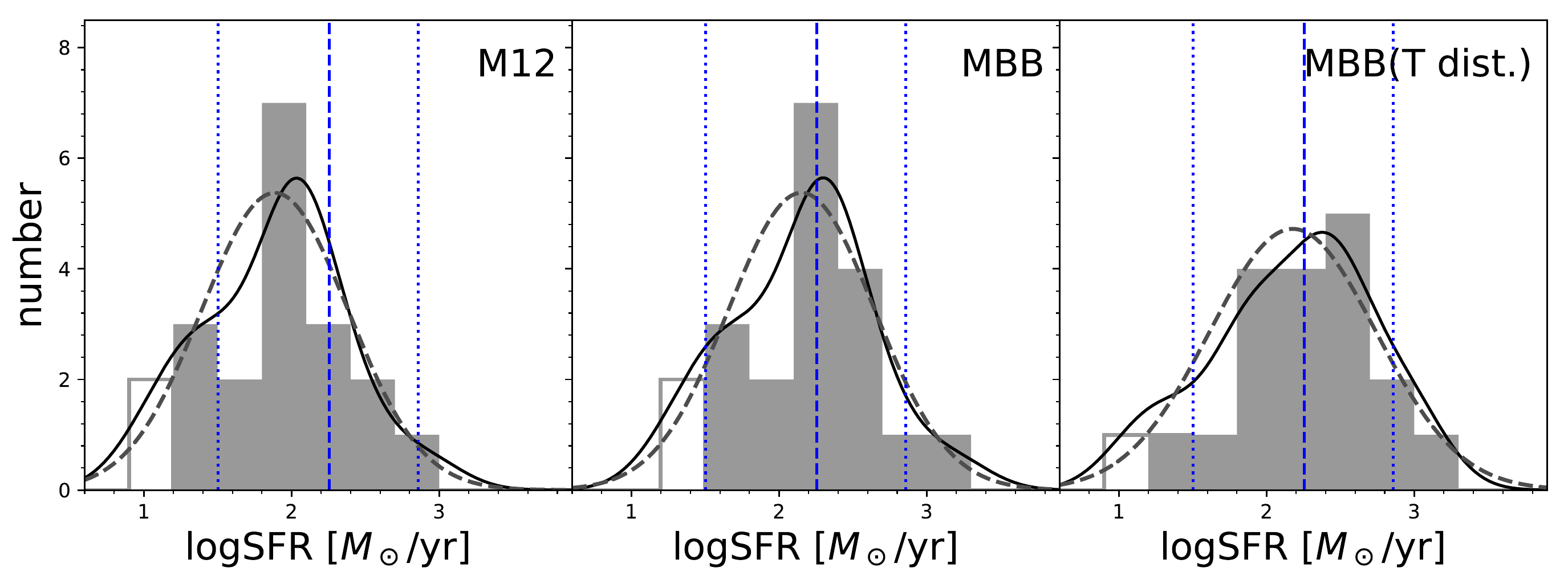} 
\caption{Observed SFR distribution for our ALMA sample (gray histogram). The black line shows a kernel density estimate of the SFR distribution, using a Gaussian kernel with 0.25~dex bandwidth and the gray dashed line represents the best fit log-normal distribution to the data as discussed in the text. Left panel: SFRs derived assuming a MS galaxy cold dust template \citep{Magdis:2012}. Middle panel: SFR based on modified blackbody with $T_{\rm d}=47$~K and $\beta=1.6$. Right panel: SFR based on modified blackbody with $\beta=1.6$ and $T_{\rm d}$ randomly drawn from a Gaussian distribution of 6~K dispersion around $T_{\rm d}=47$~K. In all panels, we indicate the location of the main sequence at $z=2$ and $\log M_\ast=11.0$ as vertical blue dashed line. The range of SFR consistent with the SF-MS, given the range of \mst expected for our sample and the dispersion of the MS of 0.3~dex, is shown as blue dotted lines. 
}
\label{fig:SFRdist} 
\end{figure*}

\rev{The second approach uses a characteristic IR SED template of MS galaxies  at $z\sim2$ as provided in \citet{Bethermin:2015}. This template is based on the SED library presented in \citet{Magdis:2012}, derived by fitting the theoretical template library of dust models by \citet{Draine:2007} to {\it Herschel} observations of distant galaxies.  In addition, we also test the template for starburst galaxies by  \citet{Bethermin:2015}.}
It has been shown that using a modified blackbody fit or SED template libraries provide generally consistent results \citep[e.g.][]{Casey:2012} for \lir.

We convert \lir\, into a SFR using the relation by \citet{Kennicutt:1998}, corrected to a \citet{Chabrier:2003} initial mass function, following SFR$/M_\odot \, {\rm yr}^{-1} = \lir / 10^{10} L_\odot$. We provide \lir\, and SFR  derived from both approaches in Table~\ref{tab:res}.

We find a tight correlation between the \lir\, estimates based on the MBB assumption and those from the \citet{Magdis:2012} MS galaxy template. However, the latter are lower by 0.27~dex, consistent with the assumption of a lower dust temperature in the MBB model of $T_{\rm d}\sim40$~K. This is consistent with \citet{Magdis:2012}, who find dust temperatures around $30-35$~K for modified blackbody fits with $\beta=1.5$ to their mean $\left< U \right>$ models. A dust temperature of 40~K is also within our adopted uncertainty on $T_{\rm d}=30-50$~K. The $z\sim2$ starburst model by \citet{Magdis:2012} is in good agreement with the MBB \lir\, values, being higher by only $0.02$~dex. Alternatively, assuming a higher dust temperature of $T_{\rm d}=60$~K would increase the \lir\, estimates by 0.43~dex.
Since several studies on luminous quasars hosts report dust temperatures closer to 47~K \citep[e.g.][]{Beelen:2006,Trakhtenbrot:2017,Duras:2017}, in the following we use the \lir\, values from the MBB model as default values, but discuss the consequences of adopting the lower \lir\, values from the MS SED template where appropriate.

\section{Results} \label{sec:discu}
\subsection{The intrinsic SFR distribution of luminous quasars} \label{sec:sfrdist}
\rev{Based on the MBB model, we find a broad range of SFRs for our sample. We list their values in Table~\ref{tab:res}. The SFRs span a range of $35-1513\, M_\odot$~yr$^{-1}$, with a median value of $180\, M_\odot$~yr$^{-1}$. The strongest star formation is detected for SDSS J1236+0500, the only object in our sample with a SFR exceeding 1000 M$_\odot$~yr$^{-1}$. }

Thanks to our high detection rate of 90\%, we are able to construct the intrinsic SFR distribution of luminous broad-line quasars at $z\sim2$ from our sample. In Figure~\ref{fig:SFRdist}, we show the SFR distribution based on the MBB model (middle panel) and the MS SED template model (left panel). For the two SF non-detections, for simplicity we assume they are located in the bin below our SFR sensitivity limit.

Assuming a single effective temperature in the modified blackbody fit for all objects is obviously a simplification, as there will be a distribution of dust temperatures. This assumption does not significantly affect the mean of the SFR distribution, but its shape. \rev{To investigate the effect of an underlying dust temperature distribution, we perform a Monte Carlo simulation. We  assign a value for $T_{\rm d}$, randomly drawn from a Gaussian distribution with mean of 47~K and a dispersion of 6~K. These values are consistent with the results by \citet{Duras:2017} for high SFR host galaxies of luminous quasars at similar redshift. For each AGN in our sample, we obtain \lir\, and SFR from a modified blackbody of this randomly drawn temperature. We derive the SFR distribution from 1000 random realizations.} The resulting SFR distribution is shown in the right panel of Figure~\ref{fig:SFRdist}. 

\begin{table}
\caption{SFR log-normal distribution parameters}
\label{tab:sfrdist}
\centering
\begin{tabular}{lcc}
\hline \hline 
Model & $<\log {\rm SFR}>$ & $\sigma_ {\rm SFR}$ \\
(1) & (2) & (3) \\
\hline 
MS galaxy template & $1.89\pm0.11$  & $0.49\pm0.08$ \\
MBB ($T_{\rm d}=47$~K) & $2.15\pm0.11$  & $0.49\pm0.08$\\
MBB ($T_{\rm d}$ distribution) & $2.14\pm0.12$  & $0.51\pm0.09$\\
\hline
\end{tabular}
\flushleft
\rev{(1) Assumed model for the SFR estimate; (2)-(3) mean and standard deviation of a log-normal fit to the SFR distribution}
\end{table}

Due to the tight correlation between the SFRs from the MBB fit and the MS galaxy template fit, the shape of the SFR distributions derived from these models are consistent, but shifted by 0.27~dex. The main difference in the random $T_{\rm d}$ case is a \rev{slight} broadening of the SFR distribution, compared to the adopted single temperature case. 

The observed SFR distribution in all three cases is fully consistent with a log-normal distribution, according to a Shapiro-Wilk test for normality. Therefore, we adopt a log-normal function to parametrize the SFR distribution, consistent with results for moderate-luminosity AGN \citep{Mullaney:2015,Scholtz:2018}.
We derive the mean, standard deviation and their confidence intervals for a log-normal SFR distribution using a Bayesian approach  \citep{Oliphant:2006}. We list the parameters of the log-normal SFR distribution for the three cases in Table~\ref{tab:sfrdist} and show it with dashed gray lines in Figure~\ref{fig:SFRdist}.

Since we do not have stellar mass estimates for our sample, but only \mbh, it is not straightforward to accurately  locate our objects in respect to the SF-MS. Nevertheless, the typical stellar mass and stellar mass distribution for luminous quasars at $z\sim2$ can be obtained from the study by \citet{Mechtley:2016}. They observed a sample of 19 luminous quasars with massive SMBHs with the {\it Hubble Space Telescope} (HST) to study the underlying host galaxy, reporting a median stellar mass of $\log \mst / M_\odot=11.1$, which we here adopt as the average \mst of our sample. For the MS relation by  \citet{Speagle:2014} at $z=2$, this mass corresponds to $\log {\rm SFR}=2.24$ for a galaxy on the MS. Adopting instead the MS relation by  \citet{Schreiber:2015} gives an almost identical result. For an expected \mst range of our sample of $\log \mst=10.5-11.5$ \citep{Mechtley:2016}, we find $\log {\rm SFR}=1.78-2.54$.
We indicate the position of the MS based on these assumptions in Figure~\ref{fig:SFRdist} as vertical blue dashed line, with the range of the MS, including a dispersion of 0.3~dex, is indicated by the blue dotted lines.

Adopting the MBB SFRs, we find the bulk of the population to be consistent with the MS of star formation. If we assume a MS SFR of $\log {\rm SFR}=2.24$ (at $\log \mst=11.1$) with dispersion of 0.3~dex, 10 quasar hosts are on the main sequence, 3 above and 7 below. Allowing for a broader range of \mst, only one object, SDSS~J1236+0500, is clearly in the starburst regime. The two SF non-detections could be located below the MS, but their upper limits are still consistent with SFRs on the MS. Interestingly, one of them is SDSS~J2313+0034, the most luminous quasar in our sample. We discuss any trends with AGN luminosity, \mbh\ and \er\ in Section~\ref{sec:agnprop}. For the case of a $T_{\rm d}$ distribution, we find largely consistent results, with an increase of the number of quasar hosts above and below the MS. For the SFR derived from the SED template, the sample mean is below the mean of the SF-MS, while the majority is still consistent with the MS within its scatter.  Five quasar host are clearly located below the MS, while only SDSS~J2313+0034 is unambiguously elevated from the MS.  
The width of the SFR distribution is broader than the typical width of the MS at this mass of $\sim0.25-0.3$~dex \citep{Speagle:2014,Schreiber:2015}. This could hint at a SFR distribution of the most luminous AGN being not consistent with the SF-MS, similar to recent results for moderate luminosity AGN \citep{Mullaney:2015,Scholtz:2018}. However, 
given the uncertainty in the stellar masses of our sample, it is possible that this broadening is a consequence of a broader underlying \mst\, distribution. 

To summarize, the SFRs of luminous $z\sim2$ quasar host galaxies are consistent with the SF-MS. Although, a fraction of the sample might be located off the MS since the SFR distribution is slightly broader than the SF MS, with possibly $5-15$\% in the starburst regime and $0-35$\% located below the MS. The latter population might be in the process of quenching star formation in the host galaxy, while the AGN is still actively accreting at a high rate. Even so, there appears to be no statistical difference between the SFR distributions of quasar hosts and typical SF galaxies of a given stellar mass. However, direct stellar mass or dynamical mass measurements will be needed to reliably pinpoint our sample with respect to the SF-MS. 

\begin{figure*}
\centering
\includegraphics[width=17cm,clip]{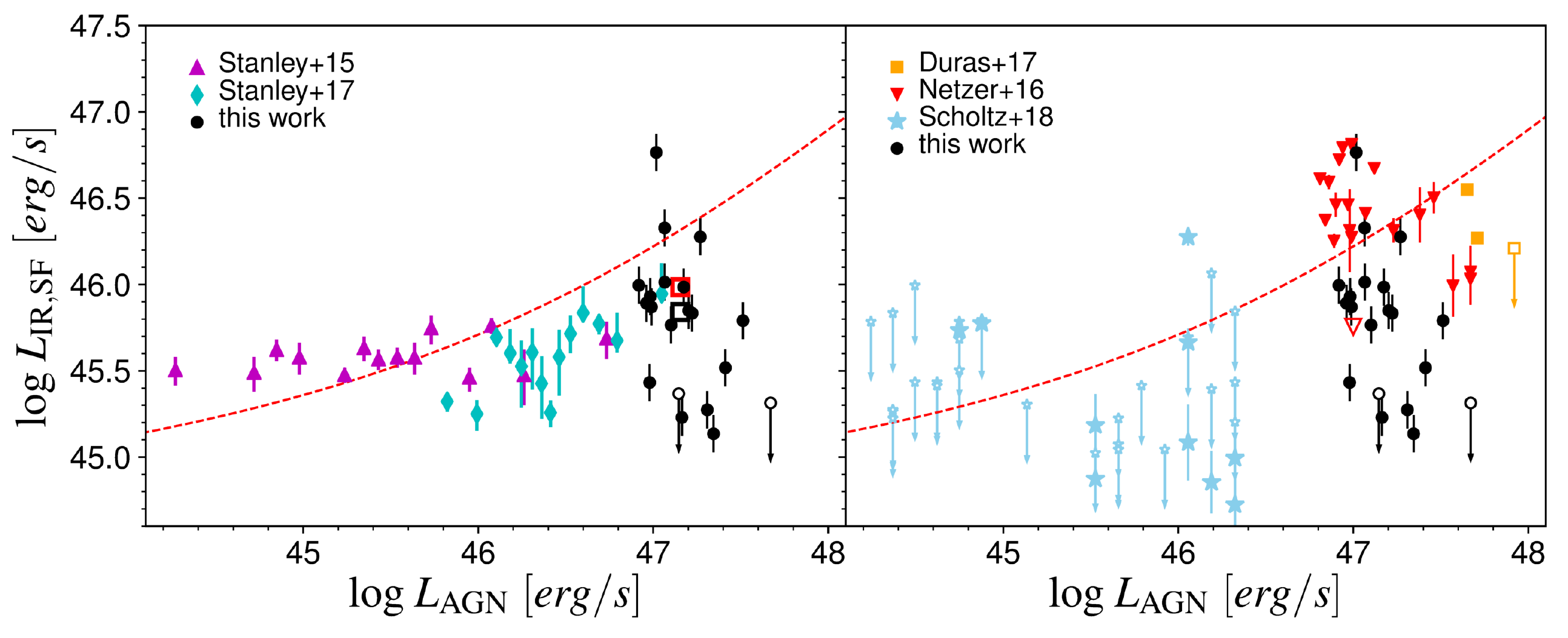} 
\caption{\lir\, due to star formation as a function of AGN $L_{\rm bol}$ for the ALMA sample studied in this work (black circles - individual measurements) in comparison to previous results in the literature. Left panel: comparison with the mean $< \lir >$ for AGN at $1.5<z<2.5$ from stacked {\it Herschel} data from \citet[][purple triangles]{Stanley:2015} and  \citet[][cyan diamonds]{Stanley:2017}. We indicate our sample median as open black square and the linear mean (corresponding to a stack) as open red square. 
Right panel: comparison to individually measured \lir\, for AGN at $1.8<z<2.5$ for moderate luminosity AGN \citep[][light blue stars]{Scholtz:2018}, luminous \citep[][red downward triangles]{Netzer:2016} and hyper-luminous quasars \citep[][orange squares]{Duras:2017}. The open symbols with arrows show upper limits and the open red triangle shows the stack for the {\it Herschel} non-detections in  \citet{Netzer:2016}. In both panels, the model by \citet{Hickox:2014} is shown by the red dashed line (see Section~\ref{sec:model}). 
}
\label{fig:lir_lagn} 
\end{figure*}

\subsection{Comparison with previous work} \label{sec:comp}
The connection between SF and AGN activity is a very active field of research, thanks to the legacy of {\it Herschel} and now ALMA. We discuss here our ALMA results for luminous $z\sim2$ quasars in the broader context of AGN-SF studies. These broadly fall into two categories 1) studies of the average SFR trends by stacking of X-ray or optically-selected AGN; and 2) studies of AGN individually detected by {\it Herschel} or in the sub-mm.

In the left panel of Figure~\ref{fig:lir_lagn} we compare our results to the average SFR trends at $1.5<z<2.0$, reported in the studies by \citet{Stanley:2015} and \citet{Stanley:2017}, based on {\it Herschel} stacks. The former work focused on moderate-luminosity X-ray selected AGN, while the latter work studied luminous optically-selected unobscured quasars from SDSS\footnote{We used the X-ray bolometric correction from \citet{Marconi:2004} to convert X-ray luminosity to bolometric luminosity in \citet{Stanley:2015}. We rescaled $L_{\rm bol}$ given in \citet{Stanley:2017} following our discussion in section~\ref{sec:mbh}.}. They consistently report an average SFR which does not show a dependence on the AGN luminosity, beyond a weak increase expected due to the different stellar mass distributions of their host galaxies. The red open square shows the linear mean for our ALMA sample, corresponding to a stacking approach. Our result is fully consistent with the highest luminosity bin by \citet{Stanley:2017}. It thus falls well into the trend discussed in that study, namely that {\it on average} AGN are hosted by normal star-forming galaxies, irrespective of their instantaneous AGN luminosity. The SFR distribution discussed in section~\ref{sec:sfrdist} agrees with this interpretation.

In contrast to the studies on stacked sources, our work in addition provides information on individual objects and on the SFR distribution. This complements recent work on moderate-luminosity AGN \citep{Mullaney:2015,Scholtz:2018}, combining observations from {\it Herschel} and ALMA, shown in the right panel of Figure~\ref{fig:lir_lagn}. In the figure we restrict the redshift range for this and all other samples shown to $1.8<z<2.5$, to approximately match our redshift range. Within this range, the sample by \citet{Scholtz:2018} includes 36 AGN with 10 of them having a SFR measurement while the rest are only upper limits. Taking into account the upper limits, there is a broad distribution of SFR, without a significant dependence on AGN luminosity. Our observations continue these trends to the highest AGN luminosities. There appears to be a mild increase in the SFR distribution, consistent with the results on stacked data. However, such an increase is expected under the assumption that AGN host galaxies are predominantly located on the SF-MS. \rev{The most luminous AGN will on average have higher \mbh\ compared to moderate-luminosity AGN. Likewise, assuming the $\mbh-\mst$ relation, they will on average have more massive host galaxies.} 

In Figure~\ref{fig:lir_lagn}, we also show SFRs measured from {\it Herschel} for individual luminous quasars over $1.8<z<2.5$ in the studies by \citet{Netzer:2016} and \citet{Duras:2017}. \rev{Here, we only show a sub-sample from these studies within the given redshift range.}
Given the sensitivity of the {\it Herschel} data their detections naturally occupy the regime of high SFR, many of them most likely located in the starburst regime. Following our discussion in  section~\ref{sec:sfrdist}, we classify AGN hosts with SFR$>690\, M_\odot/$yr as a likely starburst. In our sample only 1/20 quasars (5\%) fall into this regime, while this is the case for 9/44 \rev{(at  $1.8<z<2.5$)} in the \citet{Netzer:2016} sample (20\%).  This apparent discrepancy might be caused by systematic differences in the sample selection or in the estimation of SFR or simply due to still limited statistics. On the contrary, in the sub-mm SCUBA sample of luminous quasars by \citet{Priddey:2003} 2/34 \rev{ (6\%) of their targets within $1.8<z<2.5$ } are detected at $850\mu$m, with SFR$>1000\, M_\odot/$yr, more in line with our results. Larger samples will be required to also robustly constrain the wings of the SFR distribution. 

\begin{figure}
\centering
\resizebox{\hsize}{!}{ \includegraphics[clip]{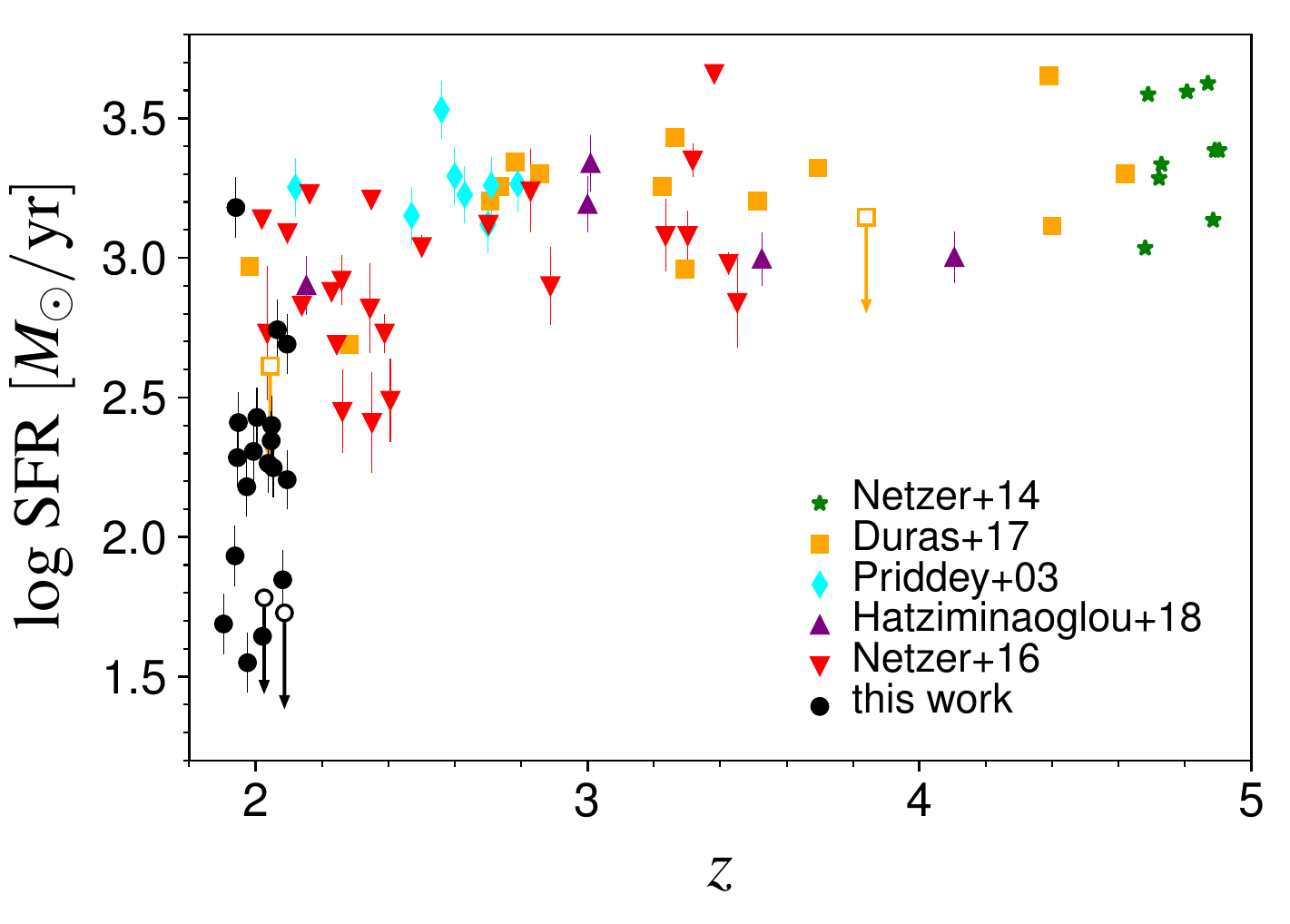} }
\caption{Individual measurements of SFR for luminous AGN with $\log L_{\rm bol}>46.9$ at $1.8<z<5$ as a function of redshift. We compare our ALMA results (black circles) to measurements based on single-dish sub-mm observations \citep[][cyan diamonds]{Priddey:2003} and the {\it Herschel} studies by \citet[][red downward triangles]{Netzer:2016}, \citet[][green stars]{Netzer:2014}, \citet[][orange squares]{Duras:2017} \rev{and \citet[purple upward triangles]{Hatziminaoglou:2018}.}
}
\label{fig:SFRvsZ} 
\end{figure}

\rev{\citet{Hatziminaoglou:2018} presented ALMA observations of 28 SDSS quasars over $2<z<4$, selected by \textit{Herschel} to be FIR bright and thus likely have extreme SFR. Their sample is very complimentary to ours. While we focus on the most luminous quasars and sample a broad range of SFR, they originally select their sample as having the highest SFRs and cover a wider range in AGN luminosity. Their study specifically probes the high SFR tail of the AGN SFR distribution, while the goal of our study is to establish the full SFR distribution at high $L_{\mathrm{bol}}$.
We show the subset of their study with comparable AGN luminosities further below.}

Recently, \citet{Falkendal:2018} presented results on the SFRs in 25 powerful high-redshift radio galaxies at $1<z<5.2$, with a median at $z\sim2.4$, using sensitive ALMA observation in combination with previous \textit{Herschel} data. Their sample of luminous AGN hosted by massive galaxies ($\log\, \mst\sim11.3$) tends to show SFR on or below the MS, with a median SFR of $110\,M_\odot/$yr, interpreted by them as representing galaxies on the way to quenching. Their sample is comparable to ours in terms of AGN luminosity and expected host stellar mass, but  while we focus on radio-quiet, unobscured AGN, they study radio-luminous, obscured AGN. Nevertheless, their results on typically low SFR in luminous AGN is consistent with our work.

In Figure~\ref{fig:SFRvsZ}, we plot individual measurements of SFR for luminous quasars, with $\log L_{\rm bol}>46.9$ over a broader range of redshift, $1.8<z<5$. \rev{We restrict the original samples to these constraints.
The samples} include the {\it Herschel} studies of luminous optically-selected quasars at $z=2-3.5$ and $z\sim4.8$ by \citet{Netzer:2016} and \citet{Netzer:2014}, respectively. We also show the study by \citet{Duras:2017} of hyper-luminous quasars from the WISSH survey \citep{Bischetti:2017} with archival {\it Herschel} detections, and the SCUBA $850\mu$m observations of a heterogenous sample of luminous quasars at $1.5<z<3$ by \citet{Priddey:2003}. For the latter study, we estimate SFR from their reported $850\mu$m fluxes in the same way as for our sample. Given the high $850\mu$m fluxes of their detections, AGN contamination is negligible for these objects. Due to their relatively low sensitivity at $850\mu$m of $\sim2.5$mJy, only quasar hosts with SFR$> 1000\, M_\odot/$yr are detected at $>3\sigma$ (in total 9/57 targets \rev{in their full sample}). 
\rev{Furthermore, we show the 5 quasars in the recent study by \citet{Hatziminaoglou:2018} with 
$L_{\mathrm{bol}}>46.9$. Again, we estimate SFR from their $850\mu m$ ALMA flux using a modified black body model, consistent to our work. 
Given their sensitivity limit, the 5 studies shown in} Figure~\ref{fig:SFRvsZ} report quasar hosts with SFR$> 300\, M_\odot/$yr, while only 3 of our targets reach as high SFR. This demonstrates how our work complements those  previous studies. While they are based on a larger parent sample and thus are more sensitive to the rare population of high SFR AGN hosts, we fill in the moderate SFR regime with individual detections, although only over a narrow range in redshift. Similarly deep ALMA studies at other redshifts will be needed to robustly assess the evolution of the SFR distribution with redshift. Potential evidence for such an evolution has been discussed in \citet{Netzer:2016}. They argue that the SFR in the most luminous quasars peaks around $z=4-5$ and declines towards lower redshift. We consider this suggestion here by comparing the fraction of quasar hosts above a certain SFR threshold, e.g. at SFR$> 1000\, M_\odot/$yr. While in our sample at $<z>=2.0$, 5\% of the sample are above this SFR cut, the fraction is 16\% in \citet{Priddey:2003} at a mean redshift of $<z>=2.3$,  17\% (12/70 with $\log L_{\rm bol}>46.9$) in \citet{Netzer:2016} at $<z>=2.7$, and 23\% (10/44) in \citet{Netzer:2014} at $<z>=4.8$. This might indicate a decline in the fraction of strong starbursts in the most luminous quasars from $z\sim5$ to $z\sim2$. Larger, homogeneously-selected quasar samples over a broad range of redshift will be required to verify this hypothesis.

Our work and those discussed above focus on optically-selected, type-1 (unobscured) AGN, i.e. heavily reddened or obscured quasars are not included in our study. The intrinsic type-2 AGN fraction is a function of luminosity, and is low at the high luminosities studied here \citep[e.g.][]{Hasinger:2008,Merloni:2014}. Since these type-2 AGN are likely unified by orientation, excluding them they will not affect our conclusions. However, reddened quasars are usually associated with a special evolutionary phase, representing young quasars in the transition phase between strongly star forming, FIR/sub-mm bright galaxies and normal unobscured quasars \citep[e.g.][]{Sanders:1988,Hopkins:2008}. Their space density, especially at the luminous end, has been suggested to be comparable to that of unobscured quasars \citep{Banerji:2015}. Sub-mm studies with ALMA found typically much higher SFR in such a reddened quasar population than what we report here \citep{Banerji:2017}, supporting the transition population hypothesis for reddened quasars. However, current sample sizes are still too small to draw firm conclusions.

\subsection{Comparison with models} \label{sec:model}
We now  discuss if  these results for the SFR distribution of luminous quasars are consistent with current models, specifically in respect to two models: 1) the phenomenological model by \citet{Hickox:2014}, and 2) results from a large cosmic volume hydrodynamical simulation \citep{Hirschmann:2014}.

\begin{figure}
\centering
\resizebox{\hsize}{!}{ \includegraphics[clip]{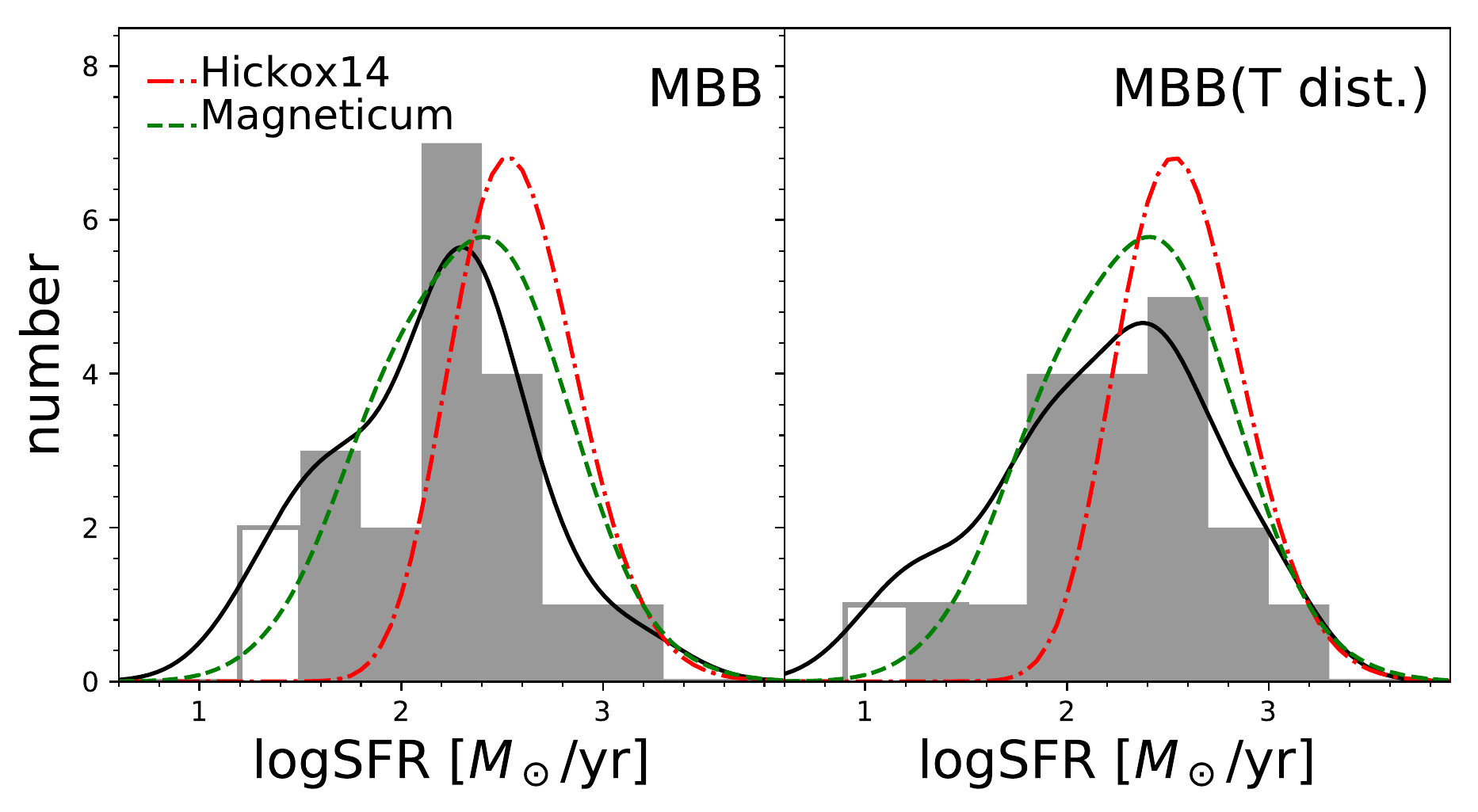} }
\caption{Observed SFR distribution, as in Figure~\ref{fig:SFRdist}, for the single dust temperature MBB model (left panel) and the random $T_{\rm d}$ MBB model (right panel). 
In addition, we compare the observations to the SFR distribution predicted from the phenomenological model by \citet[][red dashed dotted line]{Hickox:2014} and  from the large scale Magneticum cosmological simulation (green dashed line).}
\label{fig:SFRdist_comp} 
\end{figure}

\rev{The simple model by \citet{Hickox:2014} is mainly designed to simultaneously explain different observations on the AGN-SF connection. It successfully explains the flat relation of <SFR> at moderate AGN luminosities (Figure~\ref{fig:lir_lagn}), but also predicts the emergence of a tight relation at the highest $L_{\rm AGN}$, where the AGN tend to have massive SMBHs and a narrow range of \er. 
In Figure~\ref{fig:SFRdist_comp}, we show the prediction of the \citet{Hickox:2014} model for luminous AGN ($\log L_{\rm bol}>46.9$) at $z=2$ as red dashed-dotted line. Their SFR distribution has a mean of 2.58 and a dispersion of 0.31. Our results indicate a mean SFR $\sim0.4$~dex lower, suggesting a weaker correlation between SFR and AGN activity as expected in their default model. Furthermore, our results show a broader distribution compared to the \citet{Hickox:2014} model. }

With the green dashed line in Figure~\ref{fig:SFRdist_comp}, we show the predicted SFR distribution from a cosmological, hydrodynamical simulation
using the Magneticum Pathfinder simulation set \citep[Dolag et al. in prep,][]{Hirschmann:2014}, which is based on the \textit{GADGET}3 code.   Specifically, we use a large size cosmological simulation with a co-moving box size of (500Mpc)$^3$, simulated with an initial particle number of $2\times1564^3$. This box size is larger than for example the EAGLE \citep{Schaye:2015} or the Illustris \citep[e.g.][]{Vogelsberger:2014,Weinberger:2018} simulations. Therefore, in contrast to those, the Magneticum simulation enables us to probe the rare population of the most luminous AGN. This simulation is able to reproduce the AGN population up to $z=3$. In particular, at redshift $z=2$ it matches well the observed AGN luminosity function \citep{Hirschmann:2014} and the active black hole mass function \citep{Schulze:2015}. We extract SFRs and SMBH mass accretion rates from the $z=2$ snapshot of the simulation and convert the latter to $L_{\rm bol}$ assuming a radiative efficiency of $\eta=0.1$. 

Figure~\ref{fig:SFRdist_comp} shows the distribution of SFR for the 77 AGN in the simulation box with $\log L_{\rm bol}>46.9$. We find a generally good agreement between the Magneticum simulation and our results. While the mean $\log$~SFR of 2.33 is somewhat higher  and the dispersion of 0.4~dex lower than our result, a Kolmogorov-Smirnov test gives p-values of $p_{\rm KS}=0.14$ and  $p_{\rm KS}=0.46$ for the single $T_{\rm d}$ MBB model and the MBB model with $T_{\rm d}$ distribution, respectively. Thus, our sample is consistent with being drawn from the same distribution as the Magneticum simulation result, or, in other words, both the observed and simulated samples are representative of galaxies within the star-forming main sequence. While recent work demonstrated that current cosmological simulations are consistent with observations of moderate-luminosity AGN on the AGN-SF connection \citep{McAlpine:2017, Scholtz:2018}, we show that this is also the case for cosmological simulations including the most luminous AGN.

\rev{However, a larger sample of very luminous AGN both in numerical simulations and in observations are required to obtain more stringent constraints on the comparison between theory and observation at the most luminous end, especially if feedback effects are more subtle than usually assumed. We note that our work as well as the Magneticum simulation show a broader distribution that the simple \citet{Hickox:2014} model. Recently, \citet{Scholtz:2018} argued that such a broad SFR distribution could be an indication for the presence of AGN feedback, which will lead to a broadening of the SFR distribution compared to a non-AGN feedback scenario. This is based on their study of the AGN-SF connection in the Eagle simulation set, where they could compare a simulation box run with and without AGN feedback. The explored Magneticum simulation box includes an AGN feedback prescription. The agreement of our observations with this simulation may be indicative of slight effects of AGN feedback. Unfortunately, the Magneticum simulation set does not include boxes without an AGN feedback implementation that could serve as a no-feedback reference, while the Eagle simulation boxes are too small to include the luminous AGN population in significant numbers. A more detailed discussion of the full AGN-SF distribution from the Magneticum simulation suite and its comparison to observations is beyond the scope of this paper and will be presented elsewhere (Hirschmann et al., in preparation).}

\section{Conclusions} \label{sec:conclu}
\rev{
Based on ALMA Band~7 observations at $850\mu$m of 20 luminous ($\log\, L_{\rm bol}>46.9$ [erg s$^{-1}$]) unobscured quasars at $z\sim2$, our results indicate that SMBHs at high Eddington rates reside in typical star-forming galaxies without any conspicuous evidence for a relation between accretion and star formation or, in particular, with the latter being enhanced or suppressed in a systematic fashion. This brings into question the role of feedback from SMBHs in regulating the growth of the massive galaxy population. However, a picture may be emerging where star formation and accretion are supplied by gas (possibly the result of a single inflow event) from reservoirs within their host galaxy but drawn on different physical and temporal scales with each having their own efficiency with respect to gas processing.

To recap some of the details, this work expands upon previous studies of AGNs which relied on stacking \citep{Stanley:2017} of {\it Herschel} data or only FIR bright {\it Herschel-} and ALMA-detected objects \citep{Mullaney:2015}. Thanks to the high sensitivity and high spatial resolution of ALMA, we are able to detect the FIR continuum (i.e., SFR) for individual AGN for the vast majority of our targets (19/20), filling in a unique regime in the $L_{\rm AGN}-$SFR plane. We estimate the AGN contribution to the sub-mm flux by fitting AGN templates to the UV to mid-IR photometry and correct the sub-mm fluxes for this contamination. The key findings are the following.

\begin{itemize}

\item We find a broad distribution of SFR, consistent with a log-normal distribution with a mean of 140 $M_\odot$ yr$^{-1}$ and a dispersion of 0.5~dex.

\item The SFR distribution is largely consistent with that of the star-forming MS. Although, this relies on an inference of the stellar masses of their host galaxies from a related luminous quasar sample at $z\sim2$ \citep{Mechtley:2016} that is equivalent to using the $M_{BH}-M_{stellar}$ relation to infer their stellar masses.

\item The SFR distribution is both broader and shifted to lower SFR than the simple phenomenological model by \citet{Hickox:2014}, but consistent with results from a large scale cosmological hydrodynamical simulation that may hint at subtle effects of quasar feedback.

\item Comparing our results with previous work at higher redshift, we find tentative evidence for an increase of the fraction of  AGN showing intense starburst activity (SFR$>1000\, M_\odot$ yr$^{-1}$) with increasing redshift from $z\sim2$ to $z\sim5$, in line with the suggestion by \citet{Netzer:2016}. Although, we cannot make any statistically significant claims given issues with selection and sample size.

\item We do not find any statistically significant correlation between SFR and AGN properties (Section~\ref{sec:agnprop}), namely $L_{\rm bol}$, \mbh\,  and \er, \ion{C}{iv} blueshift, equivalent width and line asymmetry. However, we caution that at least part of this lack of correlation could be caused by the restriction of our sample to high luminosities, thus covering a restricted range in \mbh, and \er.

\end{itemize}

This work clearly demonstrates the unique capabilities of ALMA to determine the dust properties of luminous high-z quasars and infer their SFR distribution free of contamination from the quasar itself. It is now imperative to establish the intrinsic SFR distributions for quasars across all cosmic epochs, especially those at the earliest times \citep[$z\sim6$;][]{Izumi:2018,Izumi:2019,Shao:2019}.
}

\section*{Acknowledgements}
A.S. is supported by the EACOA fellowship. We are grateful for constructive discussions with Renyue Cen. This paper makes use of the following ALMA data: ADS/JAO.ALMA\#2017.1.00102.S. ALMA is a partnership of ESO (representing its member states), NSF (USA) and NINS (Japan), together with NRC (Canada), MOST and ASIAA (Taiwan), and KASI (Republic of Korea), in cooperation with the Republic of Chile. The Joint ALMA Observatory is operated by ESO, AUI/NRAO and NAOJ. MH acknowledges financial support from the Carlsberg Foundation via a Semper Ardens grant  (CF15-0384).




\bibliographystyle{mnras}



%
\appendix

\section{Optical spectral measurements} \label{sec:specfit}
\begin{figure*}
\centering
\includegraphics[width=17cm,clip]{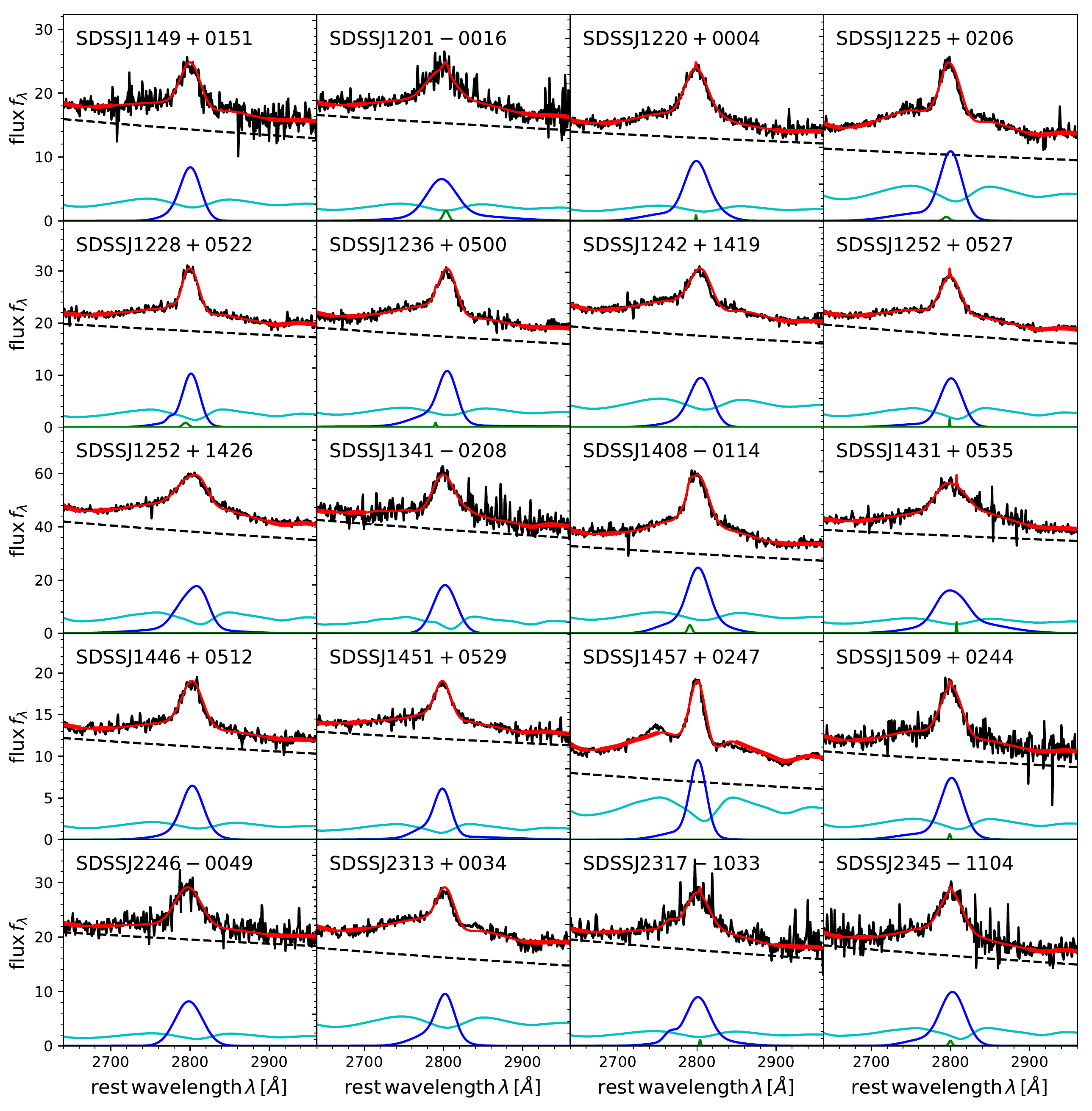} 
\caption{Optical SDSS and BOSS spectra of the \ion{Mg}{ii} emission line region. The red line shows our total best fit. We further show the best fit power law continuum (black dashed line), the \ion{Fe}{ii} component (cyan), the broad \ion{Mg}{ii} emission line (blue) and a possible weak narrow \ion{Mg}{ii} line (green).}
\label{fig:spec_m2} 
\end{figure*}

We here provide details on the fitting of the optical SDSS/BOSS spectra. We correct the spectra for galactic extinction using the extinction map from \citet{Schlegel:1998} and the reddening curve from \citet{Cardelli:1989} and shift them to their rest frame, using the \citet{Hewett:2010} redshift. We use our custom line fitting code \citep{Schulze:2018b} based on MPFIT \citep{Markwardt:2009} to model the spectral region around our target emission line. The procedure is similar to e.g. \citet{Shen:2011}. We first fit and subtract a local pseudo-continuum, consisting of a power-law and a broadened iron template. We use the UV iron template from \citet{Mejia:2016}. We do not model the Balmer continuum emission, which however will only lead to a very small overestimation of the intrinsic continuum emission at 3000\AA{} by about a factor 1.16 \citep{Trakhtenbrot:2012}. The \ion{Mg}{ii} line is fitted over the range $2700-2900$\AA{} by up to three Gaussians for the broad component and by a single Gaussian with FWHM$<900$~km s$^{-1}$ for a potential narrow component. It is not clear if a narrow component should be included in the fit for \ion{Mg}{ii}, since it is typically weak. We emphasize that for our sample its contribution is indeed marginal and does not have a significant effect on our results. The best fit results for the \ion{Mg}{ii} region are shown in Fig.~\ref{fig:spec_m2}.

In addition, we obtained a near-IR spectrum for SDSS~J2313$+$0034, the most luminous quasar in our sample, to confirm the \ion{Mg}{ii} based \mbh\ with an H$\alpha$ based estimate. We used the Nordic Optical Telescope (NOT) on August 19 2018 under good seeing condition ($\sim0.75\arcsec$) to obtain a K-band spectrum at a spectral resolution of $R=2500$. We reduced the spectrum, fit the H$\alpha$ line region and derive a \mbh\ estimate following \citet{Schulze:2017}. We find a H$\alpha$ based black hole mass of  $\log \mbh=9.72\pm0.02$, in excellent agreement with the \ion{Mg}{ii} result of $\log \mbh=9.73\pm0.04$.

\begin{figure*}
\centering
\includegraphics[width=17cm,clip]{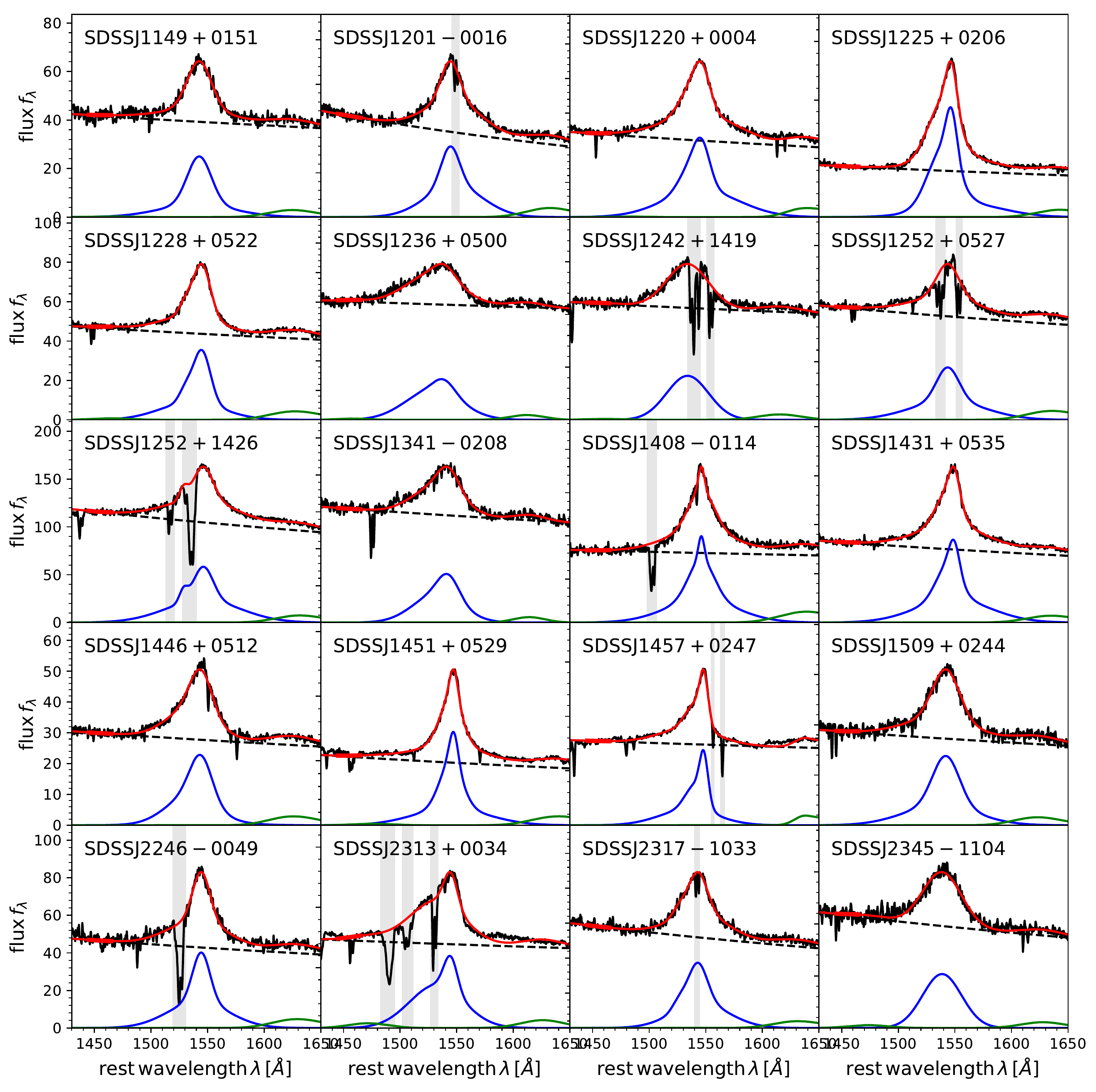} 
\caption{Optical SDSS and BOSS spectra of the \ion{C}{iv} emission line region. The red line shows our total best fit. We further show the best fit power law continuum (black dashed line), the broad \ion{C}{iv} emission line (blue) and other weak broad line contributions from \ion{He}{ii}$\lambda1640$ and \ion{O}{iii}]$\lambda1663$ (green). Gray shaded areas indicate regions that have been masked out for the fitting due to strong absorption features.}
\label{fig:spec_c4} 
\end{figure*}

For \ion{C}{iv}, we fit the local continuum by a power-law and subtract its contribution from the spectrum. We do not include an iron template since the \ion{Fe}{ii} emission around \ion{C}{iv} is typically weak.  We fit the broad \ion{C}{4} line by up to three Gaussian components over the interval $1450-1700$\AA{}. We do not attempt to model a potential narrow \ion{C}{iv} line, since if it is existent at all, its contribution is insignificant \citep[e.g.][]{Wills:1993,Vestergaard:2006}. Following \citet{Fine:2010}, we include  the \ion{He}{ii}$\lambda 1640$, \ion{O}{iii}]$\lambda1663$ and \ion{N}{iv}]$\lambda1486$ lines in the fit, modeled by a single broad Gaussian for each component, with a common line width and velocity shift. However, we do not derive any physical parameters from these fits. Several of our targets are affected by narrow absorption features on top or near the \ion{C}{iv} line. We manually mask out spectral regions affected by such features. The best fit results for the \ion{Mg}{ii} region are shown in Fig.~\ref{fig:spec_c4}.

\section{Dependence of SFR on other AGN properties} \label{sec:agnprop}
\begin{figure*}
\centering
\includegraphics[width=15cm,clip]{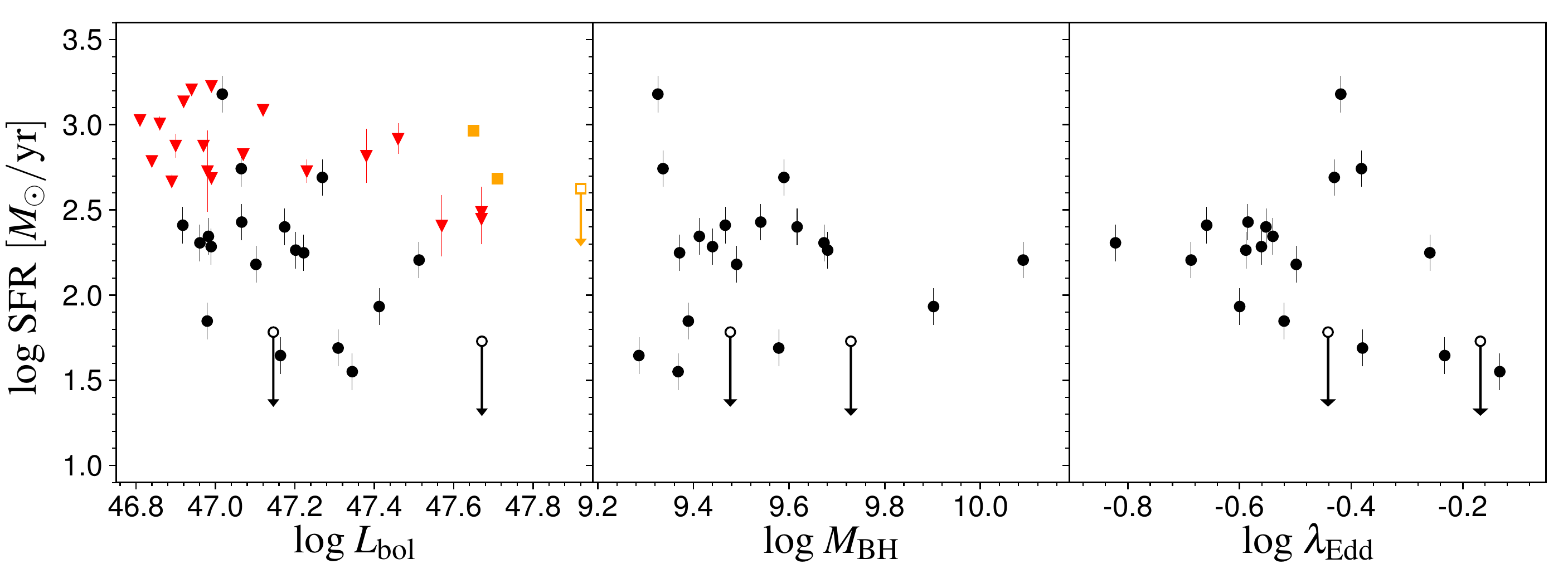} 
\caption{Dependence of SFR on $L_{\rm bol}$ (left panel), \mbh\, (middle panel)  and \er\, (right panel). In the left panel we add the SFR detections from \citet[][red downward triangles]{Netzer:2016} and \citet[][orange squares]{Duras:2017}.
}
\label{fig:SFRvsLME} 
\end{figure*}

Next, we discuss the dependence of the SFR on the AGN properties. Figure~\ref{fig:SFRvsLME} shows the distribution of SFR as a function of $L_{\rm bol}$, \mbh\,  and \er. No obvious relationship is seen for our sample. We test for the existence of a correlation via a Spearman-rank-order and Kendall's $\tau$ test, as listed in Table~\ref{tab:cortest}. There is no correlation present for \mbh\, and \er. $L_{\rm bol}$ shows tentative evidence for a weak anti-correlation. However, the high SFR detections in \citet{Netzer:2016} and \citet{Duras:2017} show that high SFR values at the highest $L_{\rm bol}$ do exist. We therefore do not consider such an anti-correlation to be significant. We note in passing that \citet{Harris:2016} also report a decline of the average SFR at their highest luminosities for $2<z<3$ quasars,  though also not at a statistically significant level.

\begin{table}
\caption{Correlation test of SFR with AGN properties}
\label{tab:cortest}
\centering
\begin{tabular}{lcccc}
\hline \hline 
Property &  $r_S$ & $p_S$ & $\tau$ & $p_\tau$ \\
(1) & (2) & (3) & (4) & (5) \\
\hline 
$\log\, L_{\rm bol}$ & $-0.48$ & 0.03 & $-0.35$ & 0.03 \\ \noalign{\smallskip}
$\log\, \mbh$ &  $-0.06$ & 0.80 & $-0.06$ & 0.70 \\ \noalign{\smallskip}
$\log\, \er$ & $-0.38$ & 0.10 & $-0.26$ & 0.10 \\ \noalign{\smallskip}
\ion{C}{iv} blueshift & $0.44$ & 0.055 & $0.35$ & 0.032 \\ \noalign{\smallskip}
EW(\ion{C}{iv}) & $-0.16$ & 0.50 & $-0.12$ & 0.48 \\ \noalign{\smallskip}
AS(\ion{C}{iv}) & $0.30$ & 0.20 & $0.23$ & 0.15 \\ \noalign{\smallskip}
\hline
\end{tabular}
\flushleft
\rev{(1) Test parameter; (2)-(3) Spearman rank order coefficient  $r_S$ and the corresponding probability $p_S$ of the null-hypothesis of no intrinsic correlation; (4)-(5)  correlation coefficient of Kendall's $\tau$ test and the corresponding probability  $p_\tau$  of the null-hypothesis.}
\end{table}

The same study also found a positive correlation between \mbh\, and SFR in their stacked sample (using \ion{C}{iv} based \mbh). However, their relation appears to flatten out above $\mbh\approx1.5\times10^9 \, M_\odot$, i.e. in the black hole mass range covered by our sample. Therefore, our results are consistent with their work. It is likely that our dynamical range in \mbh\, is too narrow to detect any potential correlation. The same is true for the Eddington ratio, where we mainly sample high \er\, quasars in our sample.

\begin{figure*}
\centering
\includegraphics[width=15cm,clip]{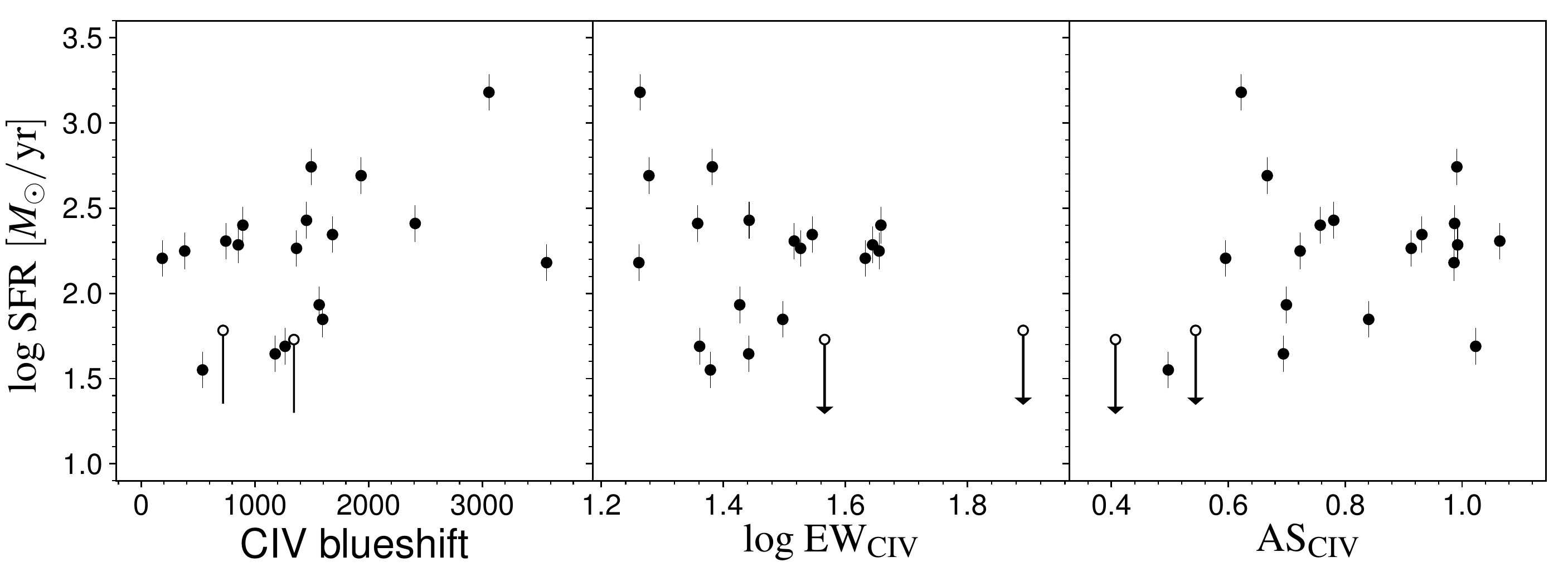} 
\caption{Dependence of SFR on \ion{C}{iv} blueshift  (left panel), \ion{C}{iv} equivalent width (middle panel)  and \ion{C}{iv} blueshift line asymmetry (right panel).
}
\label{fig:SFRvsCiv} 
\end{figure*}

In Figure~\ref{fig:SFRvsCiv}, we investigate the dependence of SFR on the properties of the \ion{C}{iv} line. The right panel shows the scatter plot with the \ion{C}{iv} blueshift. These are commonly associated with disk-wind driven outflows \citep{Gaskell:1982,Richards:2011}. We define the \ion{C}{iv} blueshift as the velocity shift of the peak of the  \ion{C}{iv} line to the peak of the  \ion{Mg}{ii} line, both derived from our best fit spectral model.  The \ion{Mg}{ii} line is known to show no systematic offset compared to the systemic redshift \citep{Shen:2016}. We find \ion{C}{iv} blueshifts in the range of zero to $\sim3500$~km/s, typical for the most luminous quasars \citep{Richards:2011,Vietri:2018}. Recently, \citet{Maddox:2017} report on average higher \ion{C}{iv} blueshift for SDSS quasars detected in the FIR by Herschel, compared to a matched sample of FIR-undetected quasars. They interpret this result as a potential signature of AGN feedback, where the AGN outflow is affecting the galaxy-scale gas content.
While the left panel of Figure~\ref{fig:SFRvsCiv} indicates a week trend of SFR with \ion{C}{iv} blueshift for our sample, this correlation is at best marginally significant. Larger samples will be required to establish such a correlation more robustly.

We do not see any other significant trend of the SFR with the properties of the  \ion{C}{iv} line, including the equivalent width (EW), line asymmetry and the prominence of narrow absorption lines. \citet{Harris:2016} report a negative correlation between  \ion{C}{iv} EW and average SFR, which they interpret as a consequence of the Baldwin effect and potentially additional factors. Our results are consistent with such an anti-correlation, however by themselves do not provide statistically significant support for it. 

The \ion{C}{iv} line in luminous quasars commonly shows blueward asymmetries in the line profile, indicative of non-virial motions, like disk-winds, outflows or non-gravitational forces \citep{Gaskell:1982, Richards:2002}.  In the right panel of Figure~\ref{fig:SFRvsCiv}, we show the relation of SFR with \ion{C}{iv} asymmetry, defined as $\rm{AS}_{CIV}=\ln \left( \frac{\lambda_{\rm red}}{\lambda_0} \right) / \ln \left( \frac{\lambda_0}{\lambda_{\rm blue}} \right)$ \citep{Shen:2012b}, ie. values lower than one indicate a blueward asymmetry, while values larger than one indicate a redward asymmetry. There is no statistically significant correlation between $\rm{AS}_{CIV}$ and SFR present for our sample, though we note that the objects with the strongest blueward asymmetry show  the lowest SFR, including the two SFR non-detections. If this blue wing component is a signature of a powerful AGN outflow, these systems could be in a state of ongoing quenching of star formation by a powerful AGN.

About 7 of the quasars show narrow absorption lines (NALs) on top of or in the vicinity of the \ion{C}{iv} line (within 5000~km s$^{-1}$). These  associated NALs are thought to be related to an AGN driven (disk-)wind \citep{Murray:1995,Elvis:2000}, affecting gas on galaxy-scales \citep[e.g.][]{Arav:2013}. Therefore, they might serve as a complimentary tracer of AGN winds. For our sample, we do not find any difference in SFR between quasars showing narrow associated absorption systems around the \ion{C}{iv} line and those without NALs close to \ion{C}{iv}, consistent with above results for other AGN wind tracers. 

\bsp	
\label{lastpage}
\end{document}